\documentclass[preprint2]{aastex}
\input epsf.sty
\usepackage{natbib}

\newcommand{\Min}{${}^{\prime}$}
\newcommand{\Sec}{${}^{\prime\prime}$}
\newcommand{\Deg}{${}^{\circ}$}

\shorttitle{Gas Kinematics of Three Hickson Compact Groups: The
Data} \shortauthors{Plana et al.}

\begin{document}


\title{Gas Kinematics of Three Hickson Compact Groups: The Data \thanks{Based on observations collected
at the European Southern Observatory, La Silla, Chile and at the
Canada-France-Hawaii Observatory, Hawaii, USA} }


\author{H. Plana \altaffilmark{1}, P. Amram \altaffilmark{2}, C. Mendes de Oliveira
\altaffilmark{3},  C. Balkowski \altaffilmark{4}, J. Boulesteix
\altaffilmark{2}}

\altaffiltext{1}{Observatorio Nacional MCT, Rua General Jose
Cristino 77, S\~ao Christ\'ov\~ao CEP: 20921-400, Rio de Janeiro,
RJ, Brazil}

\altaffiltext{2}{Laboratoire d'Astrophysique de Marseille,
Observatoire de Marseille, 2 Place Le Verrier, 13248 Marseille
Cedex 04 France}

\altaffiltext{3}{Universidade de S\~ao Paulo Instituto de
Astronomia, Geof\'{\i}sica e Ci\^encias Atmosf\'ericas,
Departamento de Astronomia, Rua do Mat\~ao 1226 - Cidade
Universit\'aria 05508-900 S\~ao Paulo SP - Brazil}

\altaffiltext{4}{Observatoire de Paris, GEPI and FRE 2459, CNRS
and Universit\'e Paris 7, 5 Place Jules Janssen, F-92195 Meudon
Cedex, France}



\begin{abstract}

We present Fabry Perot observations of three Hickson Compact
Groups, HCG 88, HCG 89 and HCG 100. We detect ionized gas in 15 group
members, three of which were previously uncatalogued objects, two
in HCG 89 and one in HCG 100. We were able to derive 2D velocity,
monochromatic and continuum maps and rotation curves for a  total
of 12 giant late-type galaxies and two dwarf galaxies.  Even with
this small sample of three groups, we can clearly see a trend of
kinematic  evolution and the different evolutionary stages of the
groups. The members of HCG 88 show almost no signs of previous or
current interactions, while HCG 100 contains at least two merging
and one strongly-interacting galaxies. HCG 89 shows members with
signs of interactions and galaxies with normal kinematics. We
therefore classify HCG 88, HCG 89 and HCG 100 respectively as an
unevolved group, a mildly interacting group and a system in the
final stage of evolution.

\end{abstract}




\section{Introduction}
Hickson Compact Groups (Hickson 1982, referred hereafter as HCGs)
are collections of three to seven galaxies inside a
three-magnitude interval, where the members have a projected
separation of the order of the galaxy diameter. The original
photometric selection criteria found spectroscopic confirmation
for 92\% of the group candidates \citep{hic89}. With a low
velocity dispersion ($\sim$ $200~km~s^{-1}$) and a high galaxy
density, Hickson Compact Groups are privileged laboratories to
study different stages of interactions, from violent interacting
systems to systems with no apparent signs of interaction.

We began in 1995 an observing program using a Fabry-Perot
instrument to obtain 2D-velocity maps of Hickson compact group
galaxies with the aim of investigating the influence of the dense
environment on the kinematics and dynamics of
their galaxy members. Such a study is the extension of a previous
effort in the same lines, in order to determine the influence of
the dense environment of galaxy clusters
\citep{amr92,amr93,amr94,amr95,amr96}. We first looked at a few
particularly interesting groups in detail, in order to understand
the evolution of galaxies in such dense environment. The groups
already studied were HCG 16 \citep{men98}, HCG 90 \citep{pla98},
HCG 18 \citep{pla00}, HCG 92 (The Stefan's
Quintet, Plana et al. 1999, Mendes de Oliveira et al. 2001) and HCG 79 (The Seyfert
Sextet, Plana et al. 2002). These studies revealed that there are large
differences in the kinematic behavior of these systems, between galaxy to
galaxy within a group and between one group to the other. Combining
the kinematics of the galaxies with other data from the literature
we were able to classify the groups in different evolutionary
stages, from evolved groups, strongly interacting, to unevolved
systems \citep{men00,amr02}.

In this paper we present Fabry-Perot data for a set of three
Hickson compact groups, HCG 88, HCG 89 and HCG 100. We were able
to derive kinematic information for 15 galaxies in these groups.
 

When available, we also present rotation curves (RCs) from previous
studies using long-slit spectroscopy (Rubin et al. 1991, hereafter RHF1991, and
Nishiura et al. 2000, hereafter NSOMT2000). A deeper comparison between our results
and these previous studies is currently under preparation in a forcoming paper
(Mendes de Oliveira et al. 2003 in prep.).

\section{Observations}

Observations were carried out in two runs at the European Southern
Observatory 3.6m telescope (ESO 3.6m) in August 1995 and at the
Canada-France-Hawaii 3.6m telescope (CFHT 3.6m) in August 1996.
During the run at the ESO 3.6m telescope, the Fabry Perot
instrument CIGALE was used. It is composed of a focal reducer
(bringing the original f/8 focal ratio of the Cassegrain focus to
f/2), a scanning Fabry-Perot and an Image Photon Counting System
(IPCS). The IPCS with a time sampling of 1/50 second and zero
readout noise makes it possible to scan the interferometer rapidly
(typically 5 seconds per channel), avoiding sky transparency,
air-mass and seeing variation problems during the exposures.

At the CFHT 3.6m, the multi-object spectrograph focal reducer
(MOS) in the Fabry-Perot mode was used, attached to the f/8
Cassegrain focus. The CCD was a STIS 2 detector, 2048 $\times$
2048 pixels with a read-out noise of 9.3 e$^{-}$ and a pixel size
on the sky of 0.86" after 2x2 binning.

Tables 1 and 2 contain the journal of observations and
observational characteristics for both runs. Reduction of the data
cubes were performed using the CIGALE/ADHOC software
\citep{bou99}.  The data reduction procedure has been extensively
described in \citep{amr92} and references therein.

Wavelength calibration was obtained by scanning the narrow Ne 6599
\AA\ line under the same conditions as the observations.
Velocities measured relative to the systemic velocity are very
accurate, with an error of a fraction of a channel width (${\rm
<~3 \,~km~s^{-1}}$) over the whole field.

Subtraction of bias, flat fielding of the data and cosmic-ray
removal have been performed for each image of the data cube for
the CFHT observations. To minimize seeing variation, each scan
image was smoothed with a gaussian function of full-width at half
maximum equal to the worse-seeing data of the data cube.
Transparency and sky foreground fluctuations have also been
corrected using field star fluxes and galaxy-free windows for the
CFHT observations.

The signal measured along the scanning sequence was separated into
two parts: (1) an almost constant level produced by the continuum
light in a narrow passband around H$\alpha $ (continuum map), and
(2) a varying part produced by the H$\alpha $ line (H$\alpha$
integrated flux map).  The continuum level was taken to be the
mean of the three faintest channels, to avoid channel noise
effects.  The H$\alpha$ integrated flux map was obtained by
integrating the monochromatic profile in each pixel. The velocity
sampling was $11~km~s^{-1}$ at CFHT and $16~km~s^{-1}$ at ESO.
Profiles were spatially binned to 3$\times$3 or 5$\times$5 pixels
in the outer parts, in order to increase the signal-to-noise
ratio. Strong OH night sky lines passing through the filters were
subtracted by determining the level of emission from extended
regions away from the galaxies \citep{lav87}.

The different kinematical parameters have been determined as followed.
The position angle (PA) of the kinematical major axis has been determined from 
the velocity field (VF) doing the average of PA estimation at a given radius from the center
. The inclination of the VF has
been determined using the method described by Amram et al. (1996), consisting of choosing the inclination 
minimizing the scattering of velocities at determined radius and at different angles from the axis.
The kinematical center also has been determined in order to get the most symmetrical rotation 
curve (RC) when both sides of the velocity plot are folded.
Morphological center corresponds to the maximum of the H$\alpha$ continuum emission.
Systemic velocities have been determined using both the velocity field and the symmetrization of the RC
(see last column of Table 4).

\section{Description of the individual groups}

In this section, we describe the main characteristics of each
individual galaxy observed. Table 3 summarizes the general
properties of the galaxies and Table 4 gives the different galaxy
parameters we derived from the continuum, monochromatic and
velocity maps. In Figures 1 to 17, we show the different maps
(continuum, monochromatic, velocity), the different rotation
curves (RCs) for each group member and, when available, comparison
with other authors' data. Group Distances have been derived using
the method described in \citet{pat97}: the mean group redshifts
come from \citet{pal93} and  have  been corrected to the centroid
of the Local Group and for infall of the Local Group towards Virgo
using an infall velocity of $170~km~s^{-1}$. The distance has been
computed using a Hubble constant $75~km~s^{-1}~Mpc^{-1}$. We found 
D  =  80.6 Mpc for HCG 88, D  =  119 Mpc for HCG 89 and D  =  71.8 Mpc for HCG 100.

\subsection{HCG 88}

\citet{hic93} described this group as a relatively loose quartet
of spiral galaxies with a very small velocity dispersion
($27~km~s^{-1}$). \citet{rib98} detects a set of six galaxies at V
 =  $6074~km~s^{-1}$ in the group region and a velocity dispersion
of $72~km~s^{-1}$. HCG 88a and b are infrared sources, emitting at
60 and 100 $\mu$m (IRAS, \citet{all96}) and HCG 88a and HCG 88d are
weak radio sources \citep{men95}. Weak CO emission has been
detected for HCG 88a (the intensity of the CO line is
$1.75~K~km~s^{-1}$, as given by \citet{bos96} and
$1.12~K~km~s^{-1}$, given by \citet{leo98}. HCG 88a shows a weak
star formation efficiency (0.22 L$_{\odot}$/M$_{\odot}$) in
comparison with the mean SFE of \citet{leo98}'s sample (0.39).
\citet{ver01} reported a H\,{\sc i} deficiency for the whole group of
\textit{$Def_{H\,{\sc I}}$} = 0.27, but individual galaxies show a rather
large spread in the H\,{\sc i} deficiency from 0.92 for HCG 88a to -0.11
for HCG 88c. \citet{pon96} gave an upper limit for the X-ray
luminosity of the group of 42.18 erg s$^{-1}$ (using the
H$_o$  =  50$~km~s^{-1}~Mpc^{-1}$).
\citet{rib98} found that this group is formed by a total of six
members and reinforce the idea that HCG 88 is an isolated compact
group.

We present continuum and monochromatic maps and velocity fields
for all four group members. \citet{shi00} detected only [N
II]-line emission and did not detect H$\alpha$-line emission in
the nucleus of HCG 88a. They classified this galaxy as an AGN. 
We detect H$\alpha$ emission throughout the disk
of HCG 88a, in a clumpy distribution. The velocity field of HCG 88a
is fairly regular but slightly twisted around the kinematic major
axis. RHF1991 and NSOMT2000  published RCs for HCG 88a, obtained
from long-slit spectroscopy, along the major axis of the galaxy.
In Fig. 3a we compare the three curves (ours, RHF1991 and
NSOMT2000). The inclination used to derive these curves is very
similar (65, 68 and 64 degrees, respectively). The RC derived by
RHF1991 matches that derived by us better than NSOMT2000 does.
Nevertheless, even in that case the agreement is not satisfactory.
The RC derived in this study is symmetric, i.e. both sides of
the curve match, which is not observed in the RC derived by
RHF1991. Indeed, the latter displays a flat approaching side and a
rising receding side, while in the present paper we find that both
sides are rising \footnote{In table 7 of RHF1991, as well as in
their Fig. 1, it is indicated that positive radial distances refer
to the tabulated position angles. Nevertheless, the two galaxies
of this group presented in their paper display an RC reversed with
respect to what we derived. We assume that this is a mistake, and
reverse the sign of their data, to compare them to ours.}. The
disagreement between both sides of RHF1991's RC is probably due to
a bad choice of the kinematic center of rotation, as it can be
suspected from the strong disagreement between both sides of their
RC around the center, within 5 arcsec.  The RC derived by
NSOMT2000 does not agree with those derived by us and by RHF1991.
Moreover, they derive an RC out to 16 arcsec only, while RHF1991
and this paper present RCs which extend out to 32 arcsec (12.5
kpc).

The SBb galaxy HCG 88b presents a very clumpy monochromatic map,
with a strong asymmetric emission in an arm-like feature in the
northeastern side and a less strong emission region in the
southern region. In contrast, on the northwestern side of the
galaxy, almost no emission can be seen.  Due to the presence of a
bar in this galaxy (SBb), the continuum image shows an asymmetric
and irregular shape, rotating from 34$^{o}$ within 10 arcsec to
160$^{o}$ at a radius of 30 arcsec. At large radius (50 arsec) the
galaxy is rather round. The velocity field shows a large scale
twist and a clear signature of a bar in the inner regions.   The
RC derived in this study is completely asymmetric,
 with a total extension of 40 arcsec
(15.6 kpc). Approaching and receding sides match well in the inner
5 arcsec and they disagree farther out. The maximum discrepancy
happens at a radius of 20 arcsec (7.8 kpc), where the approaching
side has velocities $80~km~s^{-1}$ lower than that in the receding
side.
Fig. 3b shows our data together with that from NSOMT2000. The RC
obtained by NSOMT2000 is plotted with a PA = $31^{o}$, which is
orthogonal with respect to the PA = $160^{o}$ we derived for the
major axis of this galaxy. This obviously leads to very different
results, which cannot be directly compared.  However, even if we
use NSOMT2000's kinematic parameters (inclination and PA) and
simulate a long slit in our data, we cannot reproduce their RC,
specially for the receding side (Mendes de Oliveira et al.2003, in prep.).

HCG 88c shows a rather symmetric gas distribution of what could be
a three-arm system.  The brightest $H\alpha$ knot coincides with
the continuum center.  The continuum image shows an arm-like
structure on the southeastern side of the galaxy and an asymmetric
feature (the eastern isophotes are stretched as compared to the
western isophotes).  The velocity field is regular even if a twist
can be seen (less strong than for HCG 88b).  The RC of HCG 88c,
slightly rising from 4 arcsec (1.6 kpc) to 25 arcsec (9.8 kpc),
displays a good match between the receding and approaching sides
(Fig. 4b). The RC reaches 35 arcsec (13.7 kpc, 0.9 R$_{25}$) at a
velocity of $120~km~s^{-1}$.  Fig. 6a plots the RC derived in
this study together with those derived by NSOMT2000 and RHF1991.
NSOMT2000 seems to have used a wrong PA for the major axis
(31$^{o}$ instead of 160$^{o}$ used by RHF1991).

The RC derived by RHF1991 extends to 20" (7.8 kpc) and agrees quite well to our
except the part after -20" on the receding side and the two extreme points
respectively at 24" (9.5 kpc) and at 28" (11.8 kpc) on the approaching side (see Fig. 6a).

The edge-on Sc galaxy HCG 88d shows an asymmetric continuum image
with respect to the line of nodes, due to a dust lane.  The
monochromatic image shows a clumpy structure exhibiting five
rather strong H$\alpha$ knots.  The velocity field is regular. The
RC derived in this study is symmetric and rising till 60 arcsec
(23.4 kpc) for a rotational velocity of $260~km~s^{-1}$ (Fig.
5b). The solid-body shape is probably a result of absorption
effects due to the high-inclination of the galaxy (85$^{o}$).
 The agreement between NSOMT2000 and our data (Fig. 6b)
is acceptable for the overall shape of the curves but not in the
details, specially between 10 and 20 arcsec, where NSOMT2000 find
a flat RC. Moreover, the RCs derived by NSOMT2000 ends around a
radius of 21 arcsec in the in the southwestern region and of 26
arcsec in the northeastern part. NSOMT2000 noted that HCG 88b and
88d have lower rotational velocities than
 field spiral galaxies with similar Hubble types and
luminosities; this is obviously due to the small extension of the
RC they derived.

\subsection{HCG 89}

This group is one of the most distant ones in our sample (119
Mpc). It is formed by four spiral galaxies.
\citet{all96} gave possible detections of galaxies HCG 89c and HCG 89d
and upper limits of HCG 89a and 89b, at 60 $\mu$m ( $<$ 0.18 Jy
for HCG 89a, $<$ 0.16 Jy for HCG 89b and a value of 0.31 Jy for HGC 89 cd,
 together).
CO emission has only been detected for HCG 89c \citep{leo98} with
an intensity of $0.56~K~km~s^{-1}$, which is the lowest value for
\citet{leo98}'s sample. \citet{ver01} showed that this group
exhibits almost no H\,{\sc I} deficit ($Def_{H\,{\sc I}} = 0.04$). \citet{pon96}
listed an upper limit for the X-ray luminosity
($<42.30~erg~s^{-1}$, using H$_o$  =  $50~km~s^{-1}$ Mpc$^{-1}$). 
\citet{shi00} reported nuclear emission in
several lines (H$\alpha$, [NII]6584 \AA\ and [NII]6548 \AA) for
HCG 89a and classified it as having an H\,{\sc i} activity type.

We derived maps for the four members. We also found a new
(non-catalogued) member near HCG 89a. The monochromatic map and
continuum image of HCG 89a show that the galaxy has an arm-like
structure with almost no emission in its center. A very bright
knot can be seen on the southwestern side of the disk of HCG 89a.
This was noted by RHF1991, who suspected that this knot was a
tidal dwarf candidate. The brightest part of the H$\alpha$ knot is
very round and a velocity gradient of $\sim~60~km~s^{-1}$ can be
measured from one side to the other. This very bright giant H\,{\sc ii}
region will perhaps become a tidal dwarf galaxy. However,
what we can say is that, in the present, its velocity gradient
matches perfectly the grand design velocity field of the galaxy.

At about 37 arcsec (21.3 kpc) south of HCG 89a, a bright emission
region is visible in the H$\alpha$ monochromatic map, we called HCG
89x on Fig. 7a. At very low intensity level (not shown here), our
continuum image shows that this isolated region is in fact weakly
linked (by a stellar bridge) to the main body of HCG 89a. We do
not know the absolute velocity of this object since we have no
absolute measurement of it, but only relative velocities subject
to a shift of the free spectral range of the interferometer
($265~km~s^{-1}$). The range of possible velocities is provided by
the limits of the interference filter (computed at 20 per cent of
the maximum transmission), i.e. in a range between 7870 and
$10145~km~s^{-1}$. Eight values for the systemic velocity of this
region are possible: $9190~\pm~n~\times~265~km~s^{-1}$ where
$1<n<4$). This giant H$\alpha$ knot, which has a size around 7
arcsec (4 kpc), shows a clear velocity gradient of $84~km~s^{-1}$
($\pm~42~km~s^{-1}$) decoupled from the velocity field of the main
body of the galaxy. This object is not in counter-rotation with
respect to the main body (the position of its major axis is
80$^{o}$ while the one of the main body is 54$^{o}$). Moreover, it
does not show any geometrical or velocity continuity with the main
velocity field. A fairly regular RC has been plotted for this
object (Fig. 11b). Both sides agree and reach a maximum rotational
velocity around $70~km~s^{-1}$ (for an inclination of
50$\pm$6$^{o}$). This provides a mass which is high enough for the
galaxy to be self-graviting. In conclusion, the emission-line
system south of galaxy HCG 89a is, in fact, an excellent tidal
dwarf candidate.

The kinematic and stellar major-axis position angles of HCG 89a
differ by 10$^{o}$ (see table 4).  The velocity field of HCG 89a
shows no major twisted isovelocities at large radii, which is
confirmed by the symmetric RC for radii larger than 8 arcsec (4.6
kpc). The southeastern bright extension of the velocity field is
described below.  In the inner region of the galaxy, a very steep
velocity gradient is observed to the northeast (receding side), up
to 4 arcsec (2.3 kpc, $170~km~s^{-1}$), followed by a decrease of
$40~km~s^{-1}$ to 8 arcsec (4.6 kpc, $130~km~s^{-1}$).  Farther
out, the two sides of the galaxy match.  Note that this
high-velocity spot does not present any morphological counter part
neither in the H$\alpha$ map nor in the continuum image. In the
northwestern side of the galaxy, the velocities rise regularly.
This non-bulgy galaxy is not classified as a barred galaxy. In
fact, the asymmetric signature of the velocity field could not be
due to a bar perturbation since that would give rise to a
symmetric feature. 
The RC of HCG 89a extends up to 30 arcsec (17.3
kpc) for a velocity of $215~km~s^{-1}$. RHF1991 and NSOMT2000
published RCs for this galaxy, which can be compared to that
derived here. The RC derived by RHF1991 is symmetric and reaches
25 arcsec (14.4 kpc), where the receding and approaching sides
match well except for the high velocity spot and the depression
discussed before. 
The shape of both RCs is quite similar (we also can notice the presence of the
high velocity region at -5") except for the approaching part when, our RC
decreases and their has a flat shape.
The main difference between our data and their
(Fig. 9a), comes from the velocity amplitude ($210~km~s^{-1}$ and
$130~km~s^{-1}$ respectively). This difference is due to the
kinematic inclination we deduced from the velocity field (45$^o$)
and the inclination deduced from the photometry by RHF1991
(60$^o$). Indeed, the velocity dispersion observed at each radius
for different azimuthal angles reaches a minimum for an
inclination of 45$\pm5^o$, (see \citet{amr96} for more details on
the method we used to determine the inclination).
 On the other hand, the morphological inclination is indeed the
value used by RHF1991 (60$^o$), if one excludes the southeastern
knot discussed above. As this knot is visible in the continuum
image and as we argued that this knot belongs to the main body of
the galaxy, there is no reason to exclude it to determine the
inclination of the galaxy. Including this knot, the morphological
inclination becomes compatible with the kinematic one
(45$\pm5^o$). The RC determined by NSOMT2000 is more irregular
than that of RHF1991 and ours, emphasizing the high velocity
region. In addition, it does not reach as far out as the other
RC's do (only out to 20"). The velocity amplitudes derived by
NSOMT2000 agree better with ours because the inclination
adopted (51$^o$) is more similar to the inclination derived by us.

The monochromatic and continuum images of HCG 89b show well a
spiral structure. In addition, the monochromatic image shows large
emission complexes, the largest being a region to the northeast of
the galaxy. As for HCG 89a, weak emission is visible in its
center.
The gas and stellar major-axis position angles of this galaxy
differ by 10$^o$. HCG 89b was morphologically classified as a
barred galaxy, an SBc \citep{hic93} and, indeed, a weak bar
signature can be seen through the tilt of its central
isovelocities.  The regular RC shows a continuous increase of the
rotational velocity up to $130~km~s^{-1}$ at a radius of 6 arcsec
(3.5 kpc), after which a plateau is reached. After 16 arcsec (9.2
kpc), the trend of the receding side decreases, while the
approaching side slowly rises, out to a radius of 25 arcsec (14.4
kpc), up to a velocity of $175~km~s^{-1}$. The RC derived by
RHF1991's shows a solid body rotation continuously rising from -10
arcsec to 22 arcsec. The main difference between the RC derived by
RHF1991 and that derived by us (Fig. 9b) comes from the different
values of our derived inclination and theirs ($49^o$ and $63^o$).
Even if we use our continuum map to derive a photometric
inclination (which is presumably what they used), the value we
find is 56$^o$ (see table 4), still lower than the value of
inclination used by RHF1991 to derive the RC of HCG 89b.
Simulating a long slit on our VF and using RHF1991's parameters, we can 
nevertheless derive a rotation curve matching better RHF1991 curve
(Mendes de Oliveira et al. 2003, in prep.).

A bright H$\alpha$ emission region is seen at about 38" to the
southwest of the center of HCG 89b. Its size is approximately 9
arcsec (5.2 kpc). Contrary to the giant H$\alpha$ knot seen close
to HCG 89a, this region presents a very small velocity gradient,
of the order of its velocity dispersion ($15~km~s^{-1}$). We,
therefore, cannot assess if this objet is a tidal dwarf candidate
or not. In Fig.7a, this object is labelled HCG 89x.

HCG 89c and HCG 89d are separated by only 32" and were, therefore,
imaged in one single exposure. HCG 89c is one and a half times
larger than HCG 89d (maximum radius of 24 and 15 arcsec
respectively) and it has a maximum rotational velocity almost
three times as high as that for HCG 89d (180 and $70~km~s^{-1}$
respectively).  In both cases, the agreement between both sides of
the RC is good in the central regions but diverge somewhat in the
external regions.  Both galaxies have a clumpy monochromatic map,
but HCG 89d is brighter than HCG 89c in H$\alpha$, showing clearly
three bright knots in emission.
The velocity field of HCG 89c is regular in the inner part but the
external isovelocities are peculiar. The velocity field of HCG 89d
presents an U-shape twist around the major axis which can explain
the peculiarity of the RC.  Moreover, HCG 89d shows an extension
to the south toward the direction of HCG 89c. This extension could
maybe be explained by an interaction between the two galaxies even
if the velocity difference is rather large between the extension
of HCG 89d ($8740~km~s^{-1}$) and the northern part of the velocity
field of HCG 89c ($8920~km~s^{-1}$).

\subsection{HCG 100}

This quartet, formed by four late-type galaxies, shows a bright
central Sb galaxy with strong infra-red emission. \citet{all96}
give IRAS fluxes for HCG 100a (F$_{60}\mu$m = 1.75 Jy,
F$_{100}\mu$m = 4.33 Jy) and upper limits for the remaining galaxies
in the group (F$_{60}\mu$m $<$ 0.17 Jy, F$_{100}\mu$m $<$ 0.71
Jy). Only an upper limit has been determined for the X-Ray
luminosity of this group, by \citet{pon96}. \citet{ver01} give an
H\,{\sc i} deficit \textit{$Def_{H\,{\sc I}} = 0.5$} for the whole group, higher
than the mean value for their sample. \citet{leo98} have detected
CO gas for HCG 100a only, and derived an SFE of 0.72
L$\odot$/M$\odot$, a value above the mean for their sample.
Although HCG 100d has no measured redshift, we know it is a member
of the group because we were able to detect gas emission from it.
It has, therefore, a velocity within the velocity range of our
inference filter (within $265~km~s^{-1}$). We also detected a new
point-like source (not previously catalogued) east of HCG 100b. We
could derive maps for all four members and for the new object as
well.

The monochromatic image of HCG 100a shows bipolar emission regions
but no emission is seen at the position of the continuum center,
as mentioned by \citet{vil98}. The general morphology of the
monochromatic map derived by \citet{vil98} is similar to that
derived by us. Their map reaches an extension of 31", if we
include the emission spot to the east of their map. This spot can
also be seen on our map, at the same location and distance from
the center (a radius of $\sim$ 19" along the major axis). Our
continuum map shows a regular distribution of light, with an
elliptical shape, and without any detected spiral pattern. The
velocity map shows that the extension to the east does not have a
velocity gradient but, instead, it corresponds to a single
isovelocity. We note that there is a variation of the position
angle of the major axis along the radius of the galaxy. We also
note three isovelocities which are inconsistent with the rest of
the velocity field, located northwest and southeast of the galaxy.
The RC for H100a is not symmetric, showing a flat part beyond a
radius of 8 arcsec (2.8 kpc) on the receding side and a constantly
rising velocity in the approaching side. We compare the RC with
that derived by RHF1991 (Fig. 16a). They used an inclination of
60$^o$ and a PA = 85$^o$, compared to 50$^o$ and 78$^o$ for our
data. The agreement between the two RCs is good for the receding
side but for the approaching side the velocity amplitudes given by
RFH1991 are larger than those derived by us.

HCG 100b has been classified as an Sm galaxy by \citep{hic93}, but
our data (and the velocity field in particular) suggests that this
object is more likely irregular. The extension of the
monochromatic image of this galaxy is $\sim$ 22", comparable to
what was derived by \citet{vil98}. We can clearly see
the central emission and the tail-like feature in the southeast
direction. We also note, by comparing the monochromatic and
continuum maps, that there is a difference of 25$^o$ degrees
between the PA of the two major axes. The velocity field appears
to be disturbed, with a strong variation of the position angle
along the radius. In addition, some isovelosities are inconsistent
with those measured over the whole map. The RC of HCG 100b is
obviously asymmetric: there is no match between the two sides of
the galaxy. In fact, the RC is so peculiar that it does not have
an approaching side. The rotation of this object is not consistent
with a disk-like rotator.

HCG 100c is a late-type barred spiral galaxy. The monochromatic
map shows strong central emission with an extension out to a
radius of 35". \citet{vil98} show an H$\alpha$ image of the
central part of the galaxy only. The PA of the major axis obtained
in the continuum map is slightly different from the
monochromatic map (stellar and kinematic axes differ by 20$^o$).
The velocity map clearly shows the presence of the bar, with the
central $5425~km~s^{-1}$ isovelocity. The velocity field also
shows the strong variation of the major axis along the radius and
presents inconsistent isovelocities to the northeast. The RC of
HCG 100c shows pretty good agreement between both sides out to a
radius of 13 arcsec (4.5 kpc). Further out, the receding side
continues to rise out to 14 arcsec (4.9 kpc) and then reach a
plateau. The approaching side has dropping velocities for radii
between 13 arcsec to 20 arcsec (7 kpc). This drop corresponds to
the inconsistent isovelocities to the northeast of the velocity
field. We compare the RC derived in this study with that derived
by RFH1991 (Fig. 16b). There is good agreement only in the center.

HCG 100d is a late-type edge-on spiral galaxy. No redshift
information is available for this object but, due to the narrow
FHWM of our interference filter (17 \AA\  =  $777~km~s^{-1}$), the
probability for this object to be part of the group is high. The
monochromatic image of the galaxy shows a clumpy structure,
consistent with the image produced by \citet{vil98}. The continuum
map is regular, but with a knot to the northeast. The knot
corresponds to an excess emission on the monochromatic map. The
velocity field if fairly regular but also shows a variation of the
isovelocities along the radius. The RC of HCG 100d is continuously
rising, both sides agree well within 8 arcsec (2.8 kpc) and
further out the discrepancy between both sides increases with
radius. The velocity amplitudes reach $240~km~s^{-1}$ and
$170~km~s^{-1}$ for the approaching and receding sides
respectively at a radius of 30 arcsec (10.4 kpc).

HCG 100x is an extragalactic emission object located 30" east of
HCG 100b. We did not find any mention of this object in the
literature. This object has an H$\alpha$ extension of 20" and a
very compact and intense nucleus of 5" radius. Line emission and
continuum emission are concentrated in the nucleus but we observed
a shift of 2 arcsec between both maxima. The velocity map is not
axisymmetric, as can be seen also in the RC. The approaching and
receding sides of the RC disagree completely.

\section{General discussion and Conclusions}

\subsection{A census of the interaction-related properties of the galaxies}

One of the main reasons why it is important to have the full
kinematic information for each compact-group member is to have
some insight about the possible interaction history of the galaxy.
\citet{men98} described several interaction indicators and listed
those present in the galaxies studied by them. Similarly, in Table
5, we list different indicators for all the galaxies studied here.
Different interaction scenarios, depending on the strength of the
encounter and the morphological types of the interacting systems,
will leave different signatures on the velocity fields of the
galaxies. As discussed in \citet{men98}, indicators like {\it
highly disturbed velocity field, double nuclei and double
kinematic gas component} are used to show strong evidences for
merger. Other indicators such as {\it warping, stellar and gas
major axis misalignment, tidal tail, high IR luminosity and
central activity} are used to show indications of collisions which
do not lead to merging.

\subsection {The interaction history of each galaxy}

 The data presented in this paper support the existence of regular
gaseous rotating disks in the centers of all studied galaxies,
except for HCG 100b.  Based on the interaction  indicators for
each galaxy listed in Table 5 and in the context of the various
models of galaxy interactions (\citet{bar89}, \citet{bar92,bar96})
we suggest the following histories for the 12 giant galaxies
studied here.

Galaxies HCG 88c, HCG 88d and HCG 100d seem to be undisturbed
galaxies, with no signs of interaction. However, HCG 88d and
HCG 100d may be in unfavorable inclinations to see any signs of
interaction in their velocity fields. Although we have small
number statistics, this small sample of undisturbed galaxies is
composed of the faintest members of their respective groups.

Galaxies HCG 88a, HCG 88b, HCG 89a, HCG 89b, HCG 89c and HCG 89d seem to
have suffered mild interaction. HCG 89a has a strong kinematic
peculiarity only in its center. HCG 89c has a peculiar velocity
field but no other sign of strong interaction. The remaining
galaxies show  evidence for mild interaction histories.

Galaxies HCG 100a, HCG 100b and HCG 100c are galaxies which show
strong interaction. These are all in one single group, suggesting
that in this group strong encounters have happened in the past.
This group must be, therefore, in an advanced stage of evolution.
HCG 100a may have also suffered a major accretion event, given its
very peculiar velocity field.

\subsection {The evolutionary stage of each group}

Based on the description above, our summary for the evolutionary
history of the three groups is the following.

HCG 88 does not display many indicators of having suffered
interaction and it displays no indicators of merging. HCG 88c and
HCG 88d do not show any interaction signature at all. We should
note, however, that the RC for HCG 88d is difficult to obtain
because of the high inclination of the disk. HCG 88b has the most
disturbed velocity field for this group and it has the least
regular RC of the four galaxies of the group. However, it
cannot be classified as ``highly disturbed'', as other
galaxies in group HCG 100 are. HCG 88a is an active galaxy but
there is no evidence in its velocity field, that this galaxy has
suffered interactions in the recent past. This is not an
unexpected result, since it is well known that although active
galaxies often appear to be in interaction, it is not always the
case. In conclusion, HCG 88 may be an unevolved group, which has
recently formed, with low level of interaction. This is in
agreement with the conclusion from the H\,{\sc i} kinematic study of this
group \citep{ver01}.

The list of Table 5 shows that the galaxies in HCG 89 have more
positive indicators than those in HCG 88. HCG 89c seems to have
the most disturbed velocity field. HCG 89b, HCG 89c and HCG 89d
show mildly disturbed rotation curves. HCG 89a has a large
disturbance in its center, perhaps because of the canibalism of a
small galaxy, or perhaps due to an interaction with the bright
emission blob present in the eastern part of the galaxy. The
difference of systemic velocities between HCG 89c and HCG 89d is
only $40~km~s^{-1}$ and the projected distance is 30 arcsec. It is
tempting to say that these two objects may be in interaction. The
velocity field of HCG 89d shows an extension to the south, towards
HCG 89c, but the velocity difference between the two galaxies is
$180~km~s^{-1}$, which is too large for a scenario of gas exchange
between the two objects. This distant group does not have a lot of
ISM material but HCG 89d clearly shows an excess of warm gas in
comparison with HCG 89c. We conclude that this group is in an
intermediate evolutionary phase, not as unevolved as HCG 88 but
also not as evolved as HCG 100.

HCG 100 shows many signs of interactions. Three of the four giant
galaxies of the group have a highly disturbed velocity field and
no member has a RC which is completely regular and axisymetric
(however we should note that the peculiarity for HCG 100d is small
but difficult to detect due to its high inclination). Change of PA
of the major axis along the radius is also present for three
galaxies in this group. However, no central activity has been
found for any group member. The most disturbed group member is HCG
100b. This object may be the product of a merger.
In conclusion, HCG 100 is the most evolved group of the three
studied here.

\acknowledgments

H.P. would like to thank the Brazilian PRONEX program, for
financial help to attend the conference ``Galaxies: The Third
Dimension" in Cozumel (Mexico), Dec. 3-7 2001 and the Marseille
Observatory Scientific Council for financial help during his stay
at the Marseille Observatory in May/June 2002. H.P. also
acknowledges the financial support of Brazilian Cnpq under
contract 150089/98-8. CMdO acknowledges financial support from
PRONEX and FAPESP. The authors thank J.L. Gach for helping during
the observations. The NASA/IPAC Extragalactic Database (NED) is
operated by the Jet Propulsion Laboratory, California Institute of
Technology, under contract with NASA.

\clearpage

%
%



\begin{figure*}

\figurenum{1a} \vspace{-2.5cm}

\epsfxsize=17cm \epsfbox{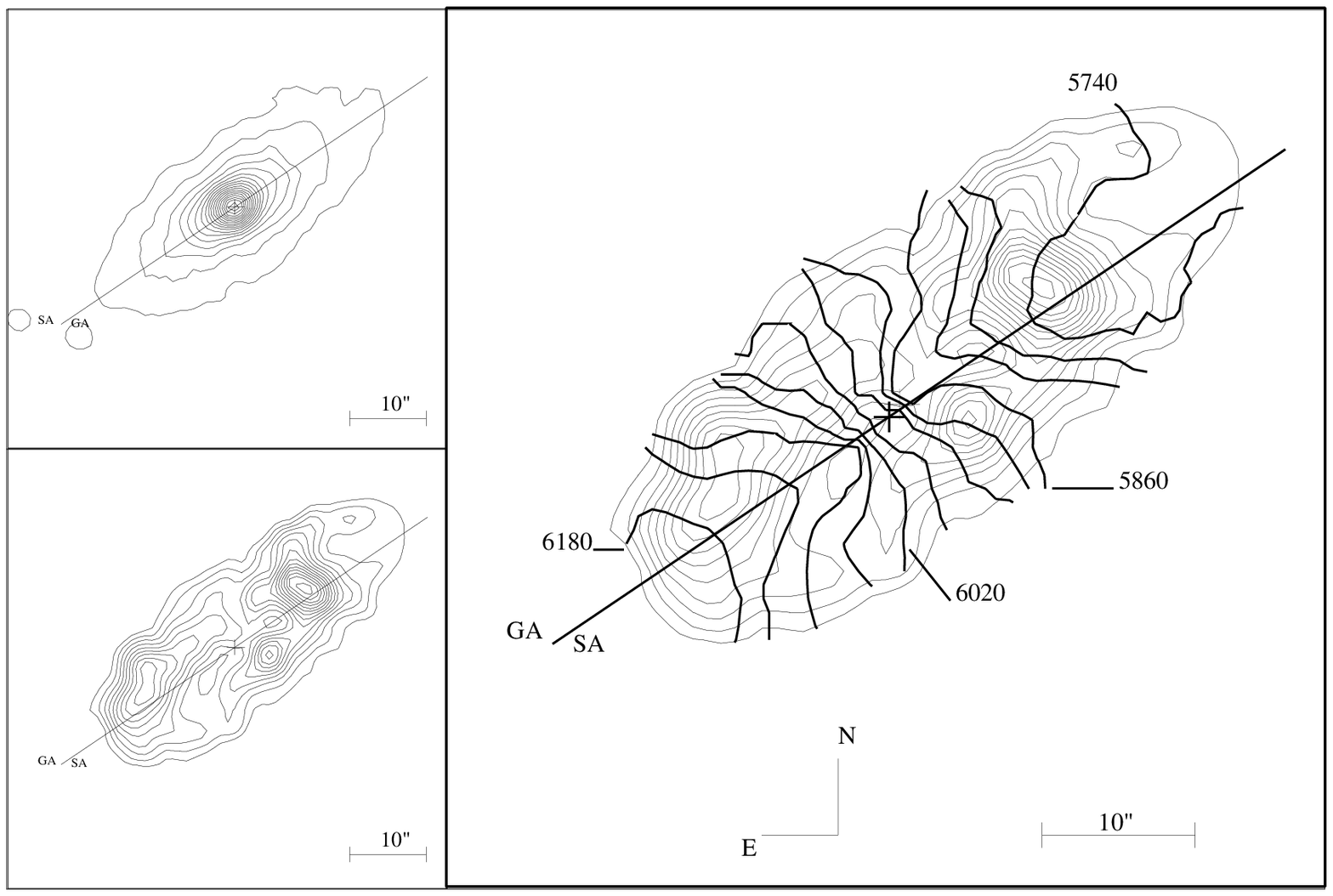}

\caption{HCG 88a: Upper left
panel represents the continuum map with respectively the stellar
axis (SA) and the monochromatic map axis (GA). Lower left panel is
the monochromatic map. Right panel is the velocity field (in
bold) superimposed to the monochromatic map. GA axis represents
the major axis of the velocity field, SA axis is the major axis of
the continuum map. Bold cross represents the kinematical center.}

\figurenum{1b}

\epsfxsize=15cm \epsfbox{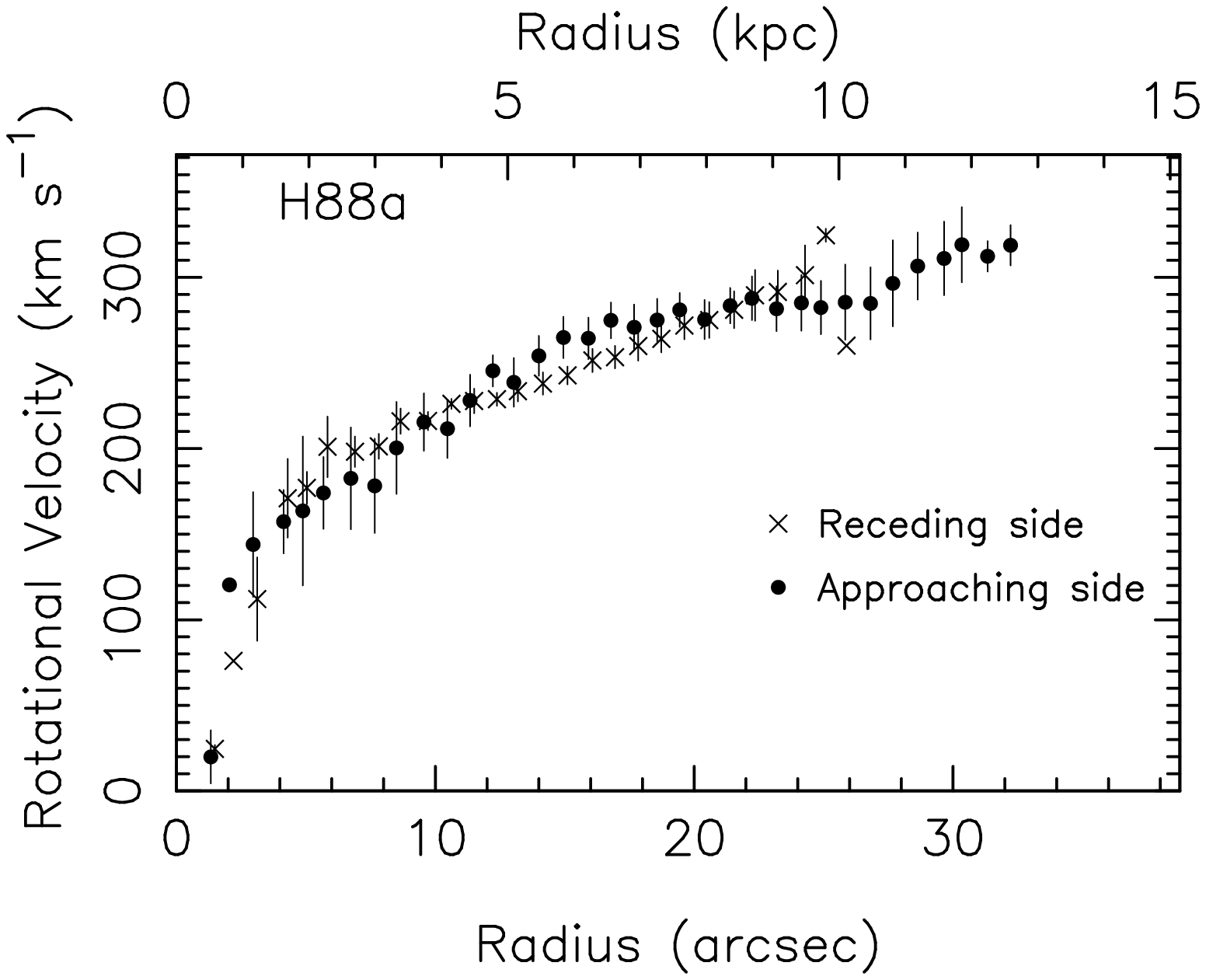}

\caption{HCG 88a: Rotation curve. Arrow represents the R$_{25}$
isophote from RC3 catalog, when available.}

\end{figure*}

\clearpage


\begin{figure*}

\figurenum{2a} \vspace{-2.5cm}

\epsfxsize=17cm \epsfbox{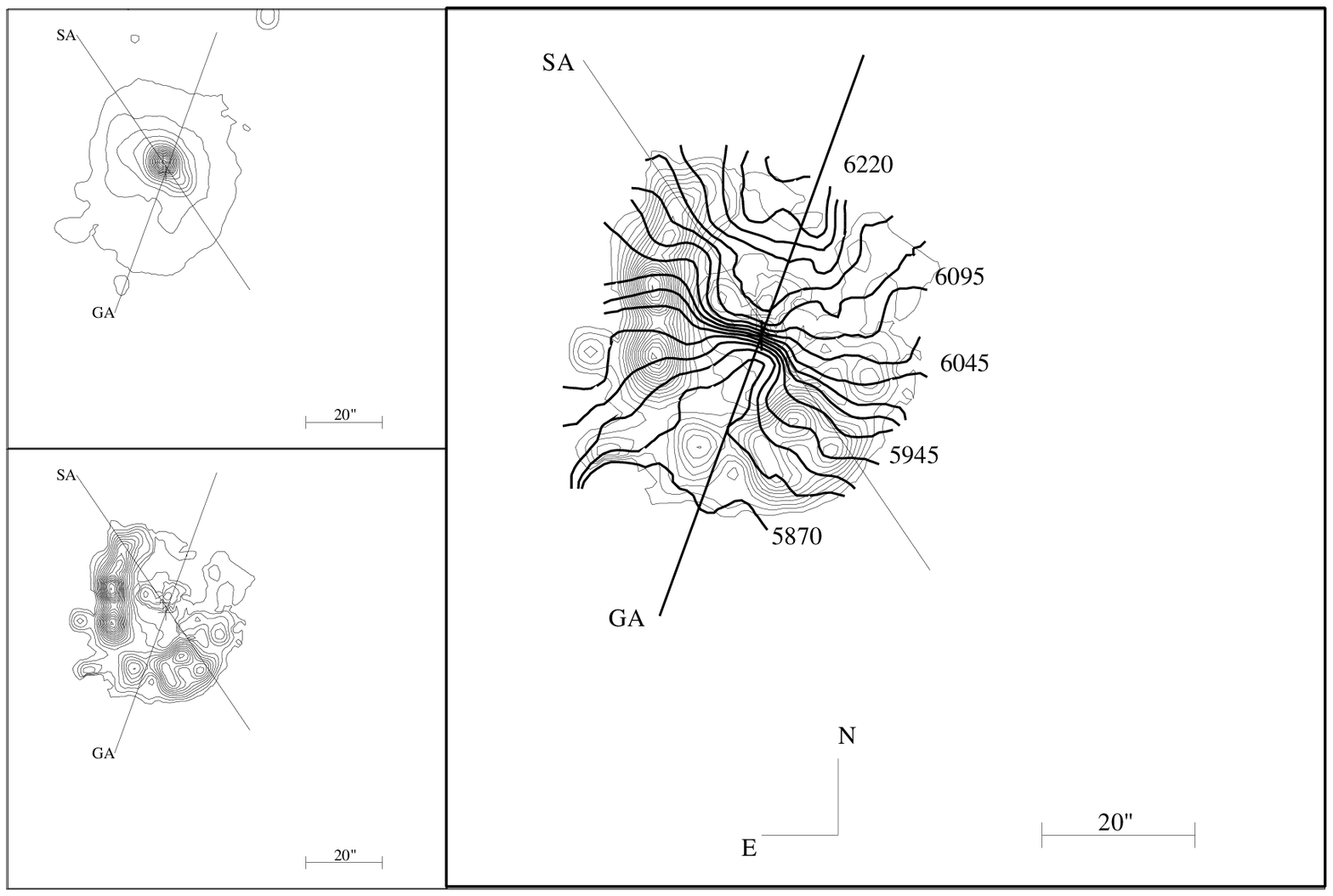}

\caption{HCG 88b: same as Fig.1a}

\figurenum{2b}

\epsfxsize=15cm \epsfbox{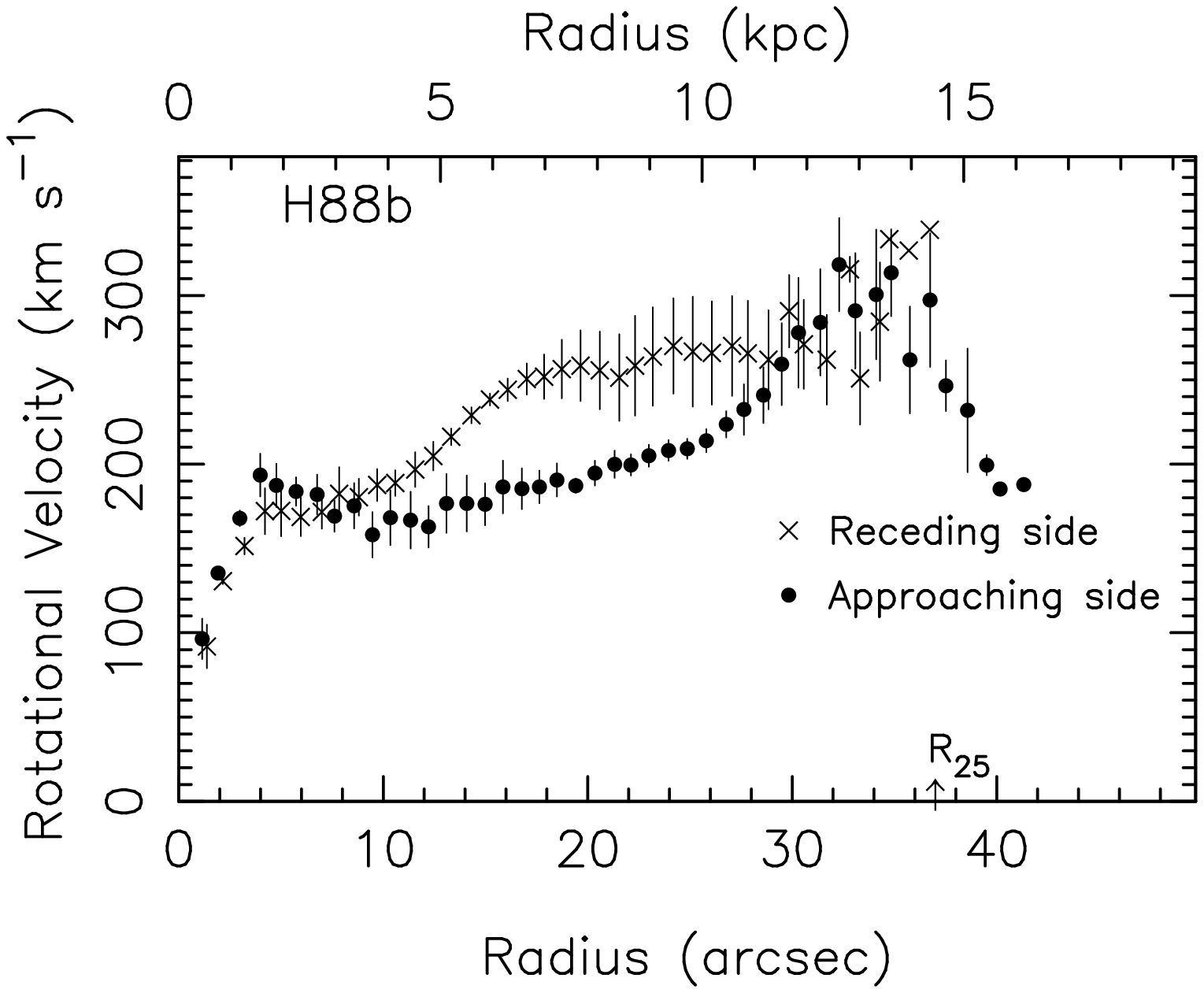}

\caption{HCG 88b: same as Fig.1b}

\end{figure*}

\clearpage


\begin{figure*}

\figurenum{3a} \vspace{-2.5cm}

\epsfxsize=15cm \epsfbox{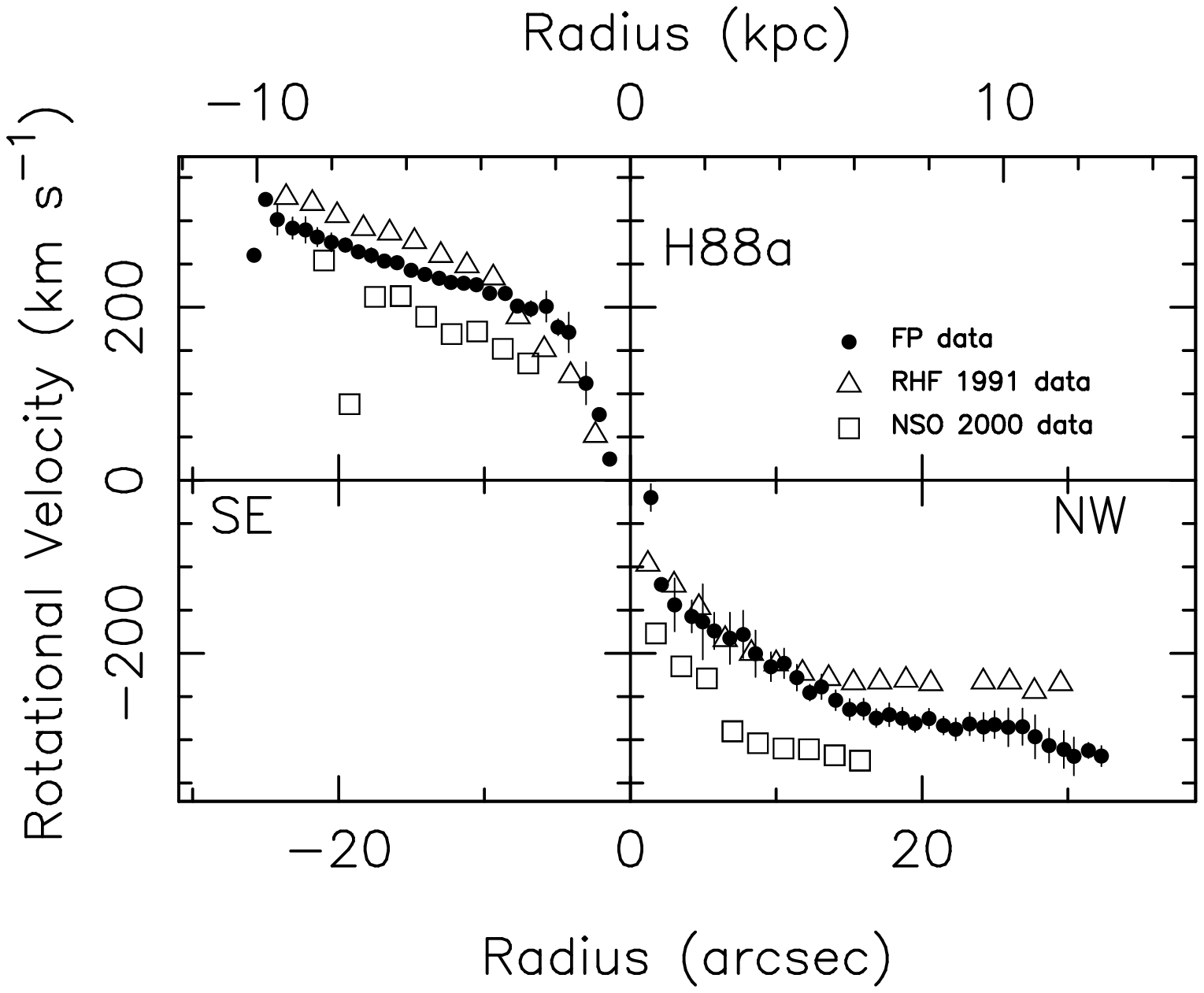}

\caption{HCG 88a: we are presenting different rotation curves using our data, 
RHF1991
data and NSOMT2000 data. The kinematical parameters used to plot
the rotation curves are, for the inclination, $65^o$, $68^o$ and $64^o$
respectively and for the position angle of the major axis,
$128^o$, $127^o$ and $132^o$ respectively.}

\figurenum{3b}

\epsfxsize=15cm \epsfbox{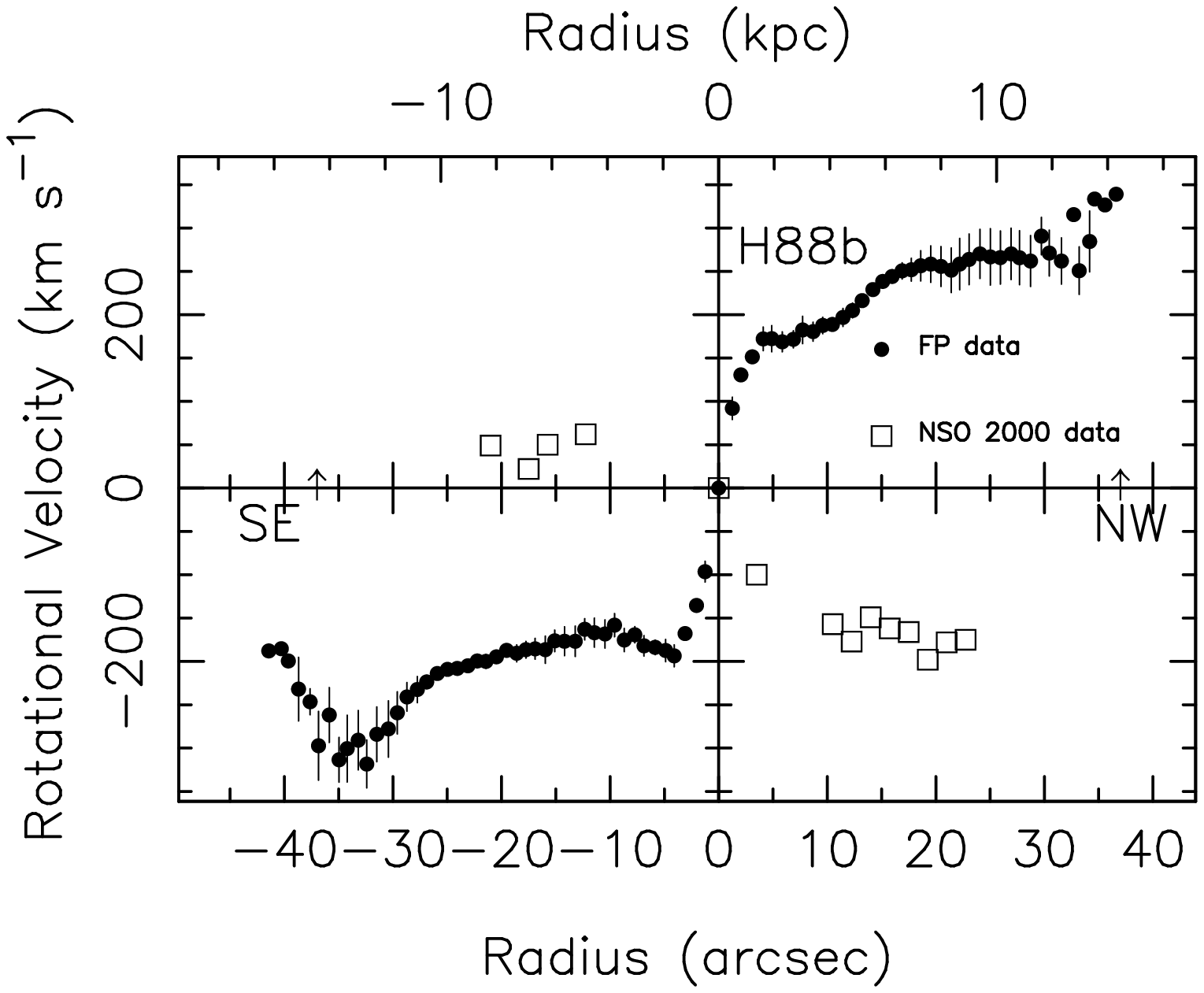}

\caption{HCG 88b:  we are presenting different rotation curves using our data and
NSOMT2000 data. The kinematical parameters used to plot the rotation curves
are, for the inclination, $46^o$ and $43^o$ respectively and for
the position angle of the major axis, $160^o$ and $31^o$
respectively.}

\end{figure*}

\clearpage


\begin{figure*}

\figurenum{4a} \vspace{-2.5cm}

\epsfxsize=17cm \epsfbox{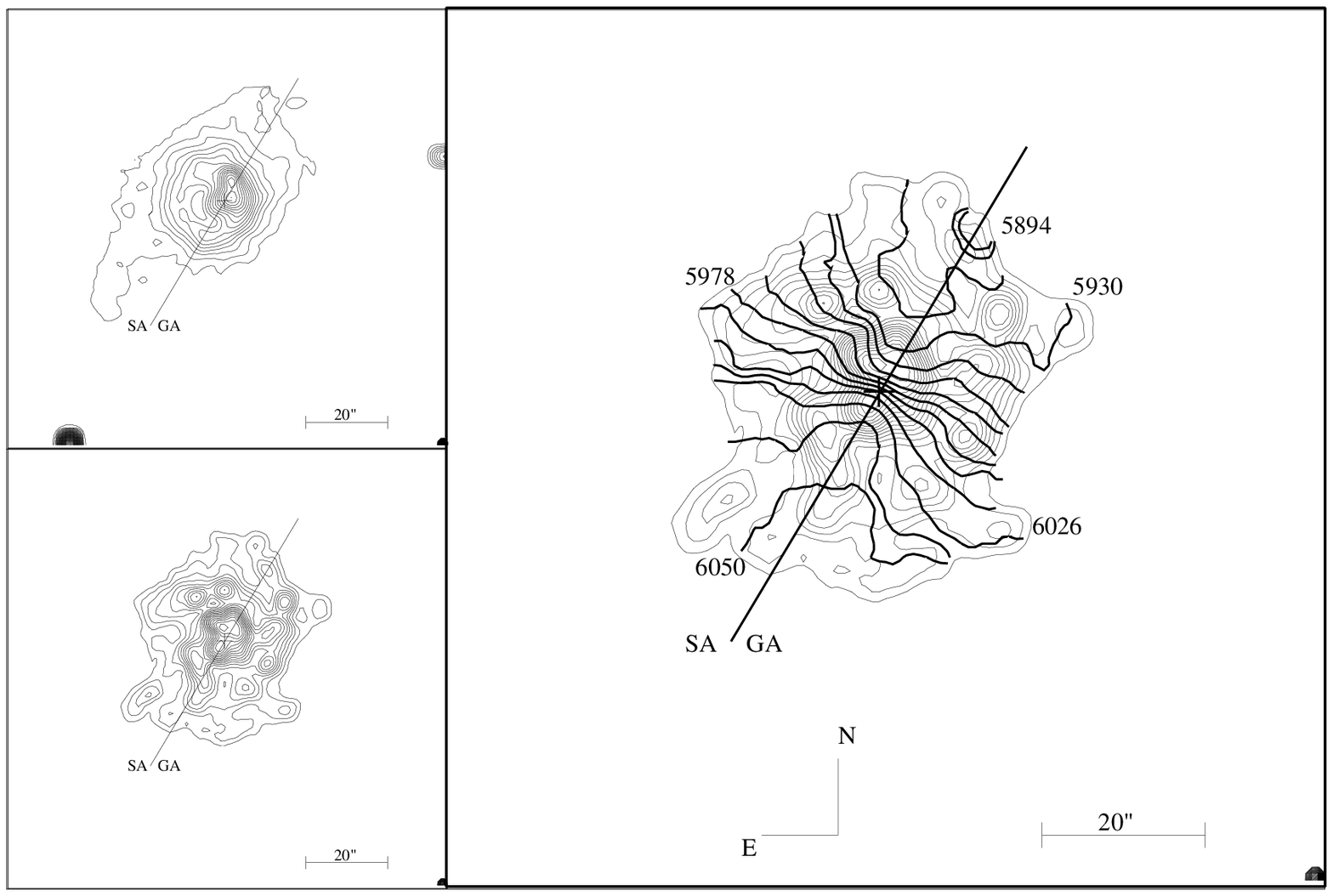}

\caption{HCG 88c: same as Fig.1a}

\figurenum{4b}

\epsfxsize=15cm \epsfbox{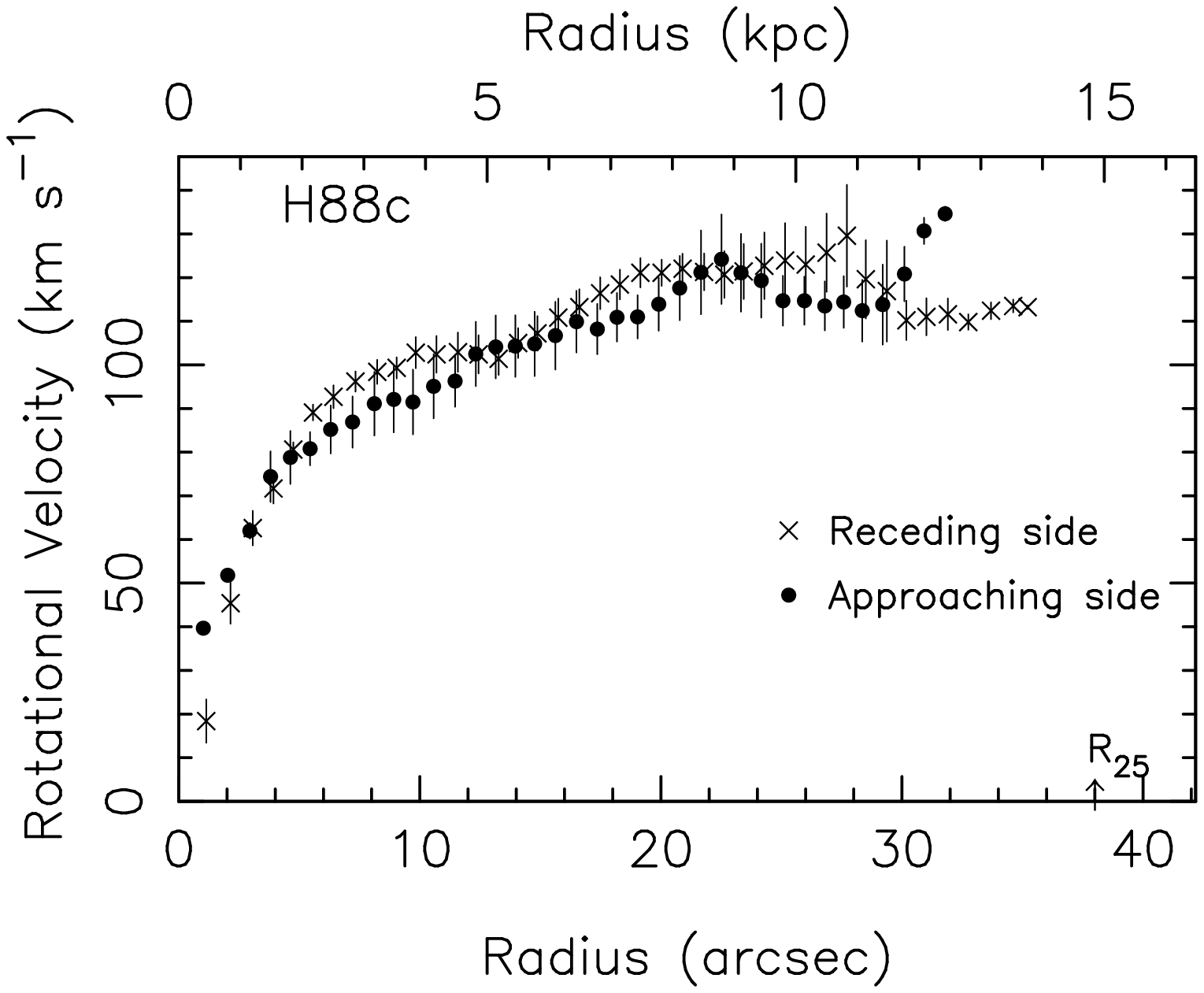}

\caption{HCG 88c: same as Fig.1b}

\end{figure*}

\clearpage


\begin{figure*}

\figurenum{5a} \vspace{-2.5cm}

\epsfxsize=17cm \epsfbox{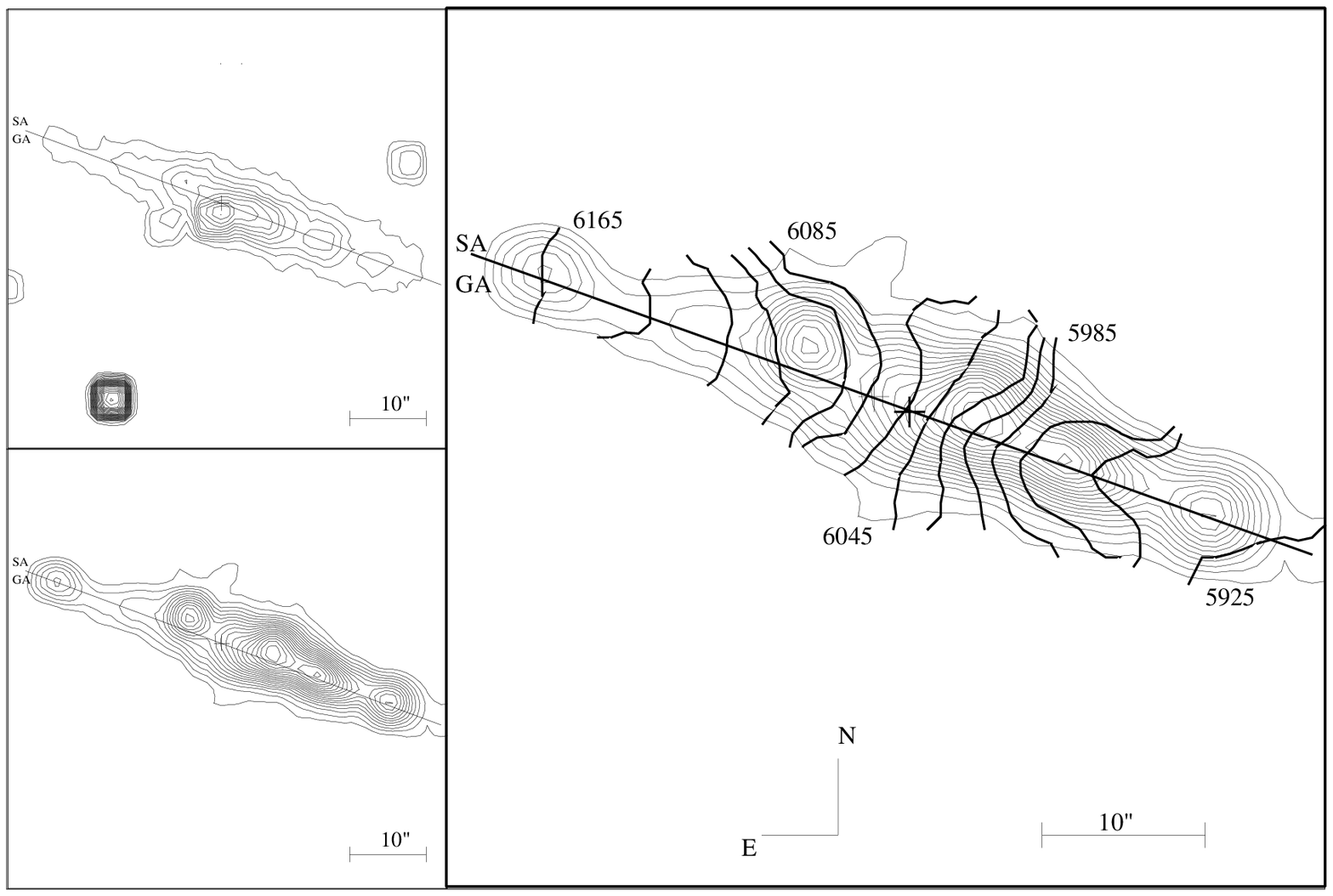}

\caption{HCG 88d: same as Fig.1a}

\figurenum{5b}

\epsfxsize=15cm \epsfbox{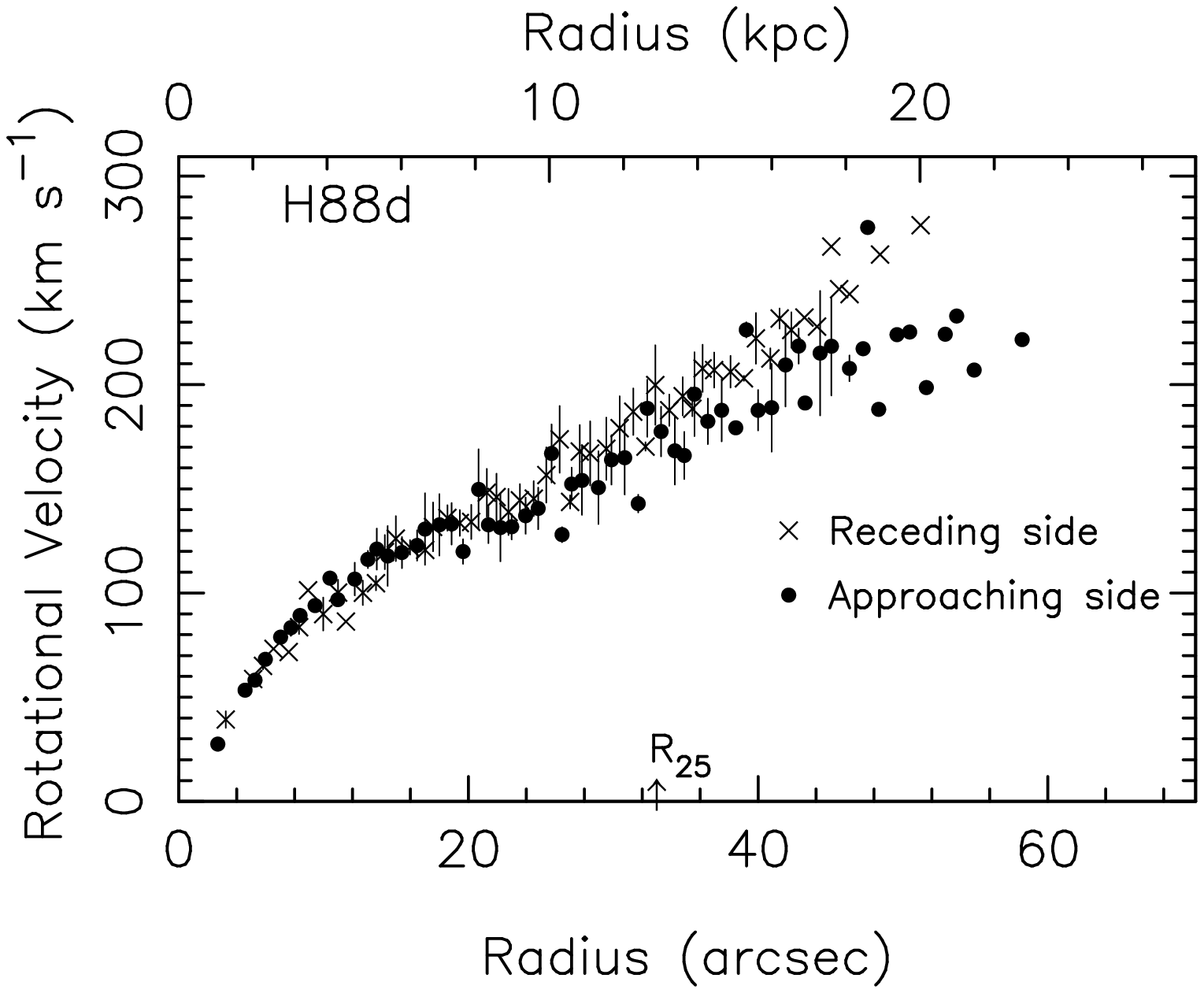}

\caption{HCG 88d: same as Fig.1b}

\end{figure*}

\clearpage


\begin{figure*}

\figurenum{6a} \vspace{-2.5cm}

\epsfxsize=15cm \epsfbox{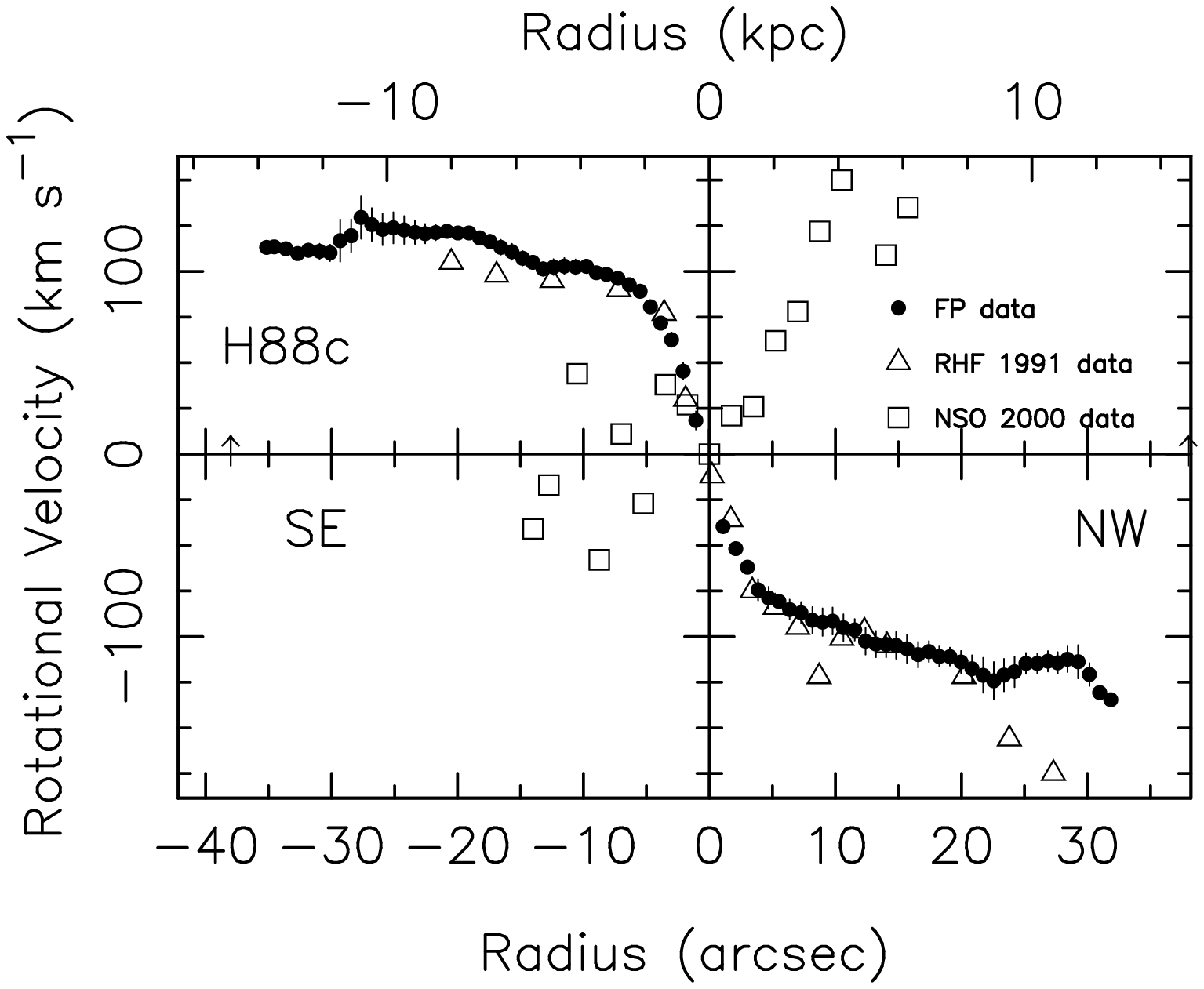}

\caption{HCG 88c: we are presenting  different rotation curves using our data,
RFH1991 data and NSOMT2000 data. The kinematical parameters used
to plot the rotation curves are, for the inclination, $42^o$, $34^o$ and
$32^o$ respectively and for the position angle of the major axis,
$150^o$, $160^o$ and $31^o$ respectively.}

\figurenum{6b}

\epsfxsize=15cm \epsfbox{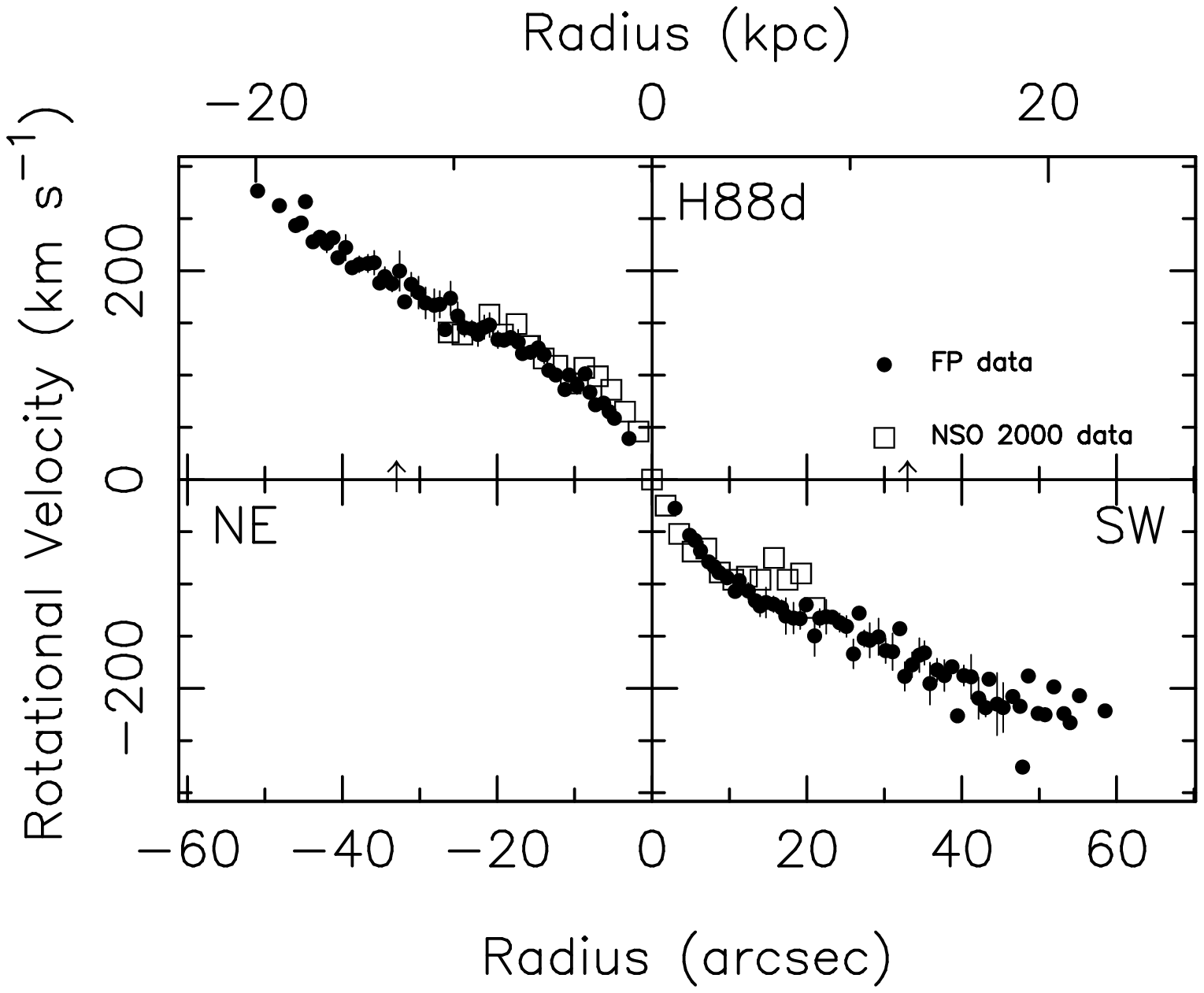}

\caption{HCG 88d: we are presenting  different rotation curves using our data and
NSOMT2000 data. The kinematical parameters used to plot the rotation curves
are, for the inclination, $85^o$ and $90^o$ respectively and for
the position angle of the major axis, $70^o$ and $71^o$
respectively.}

\end{figure*}

\clearpage



\begin{figure*}

\figurenum{7a} \vspace{-2.5cm}

\epsfxsize=17cm \epsfbox{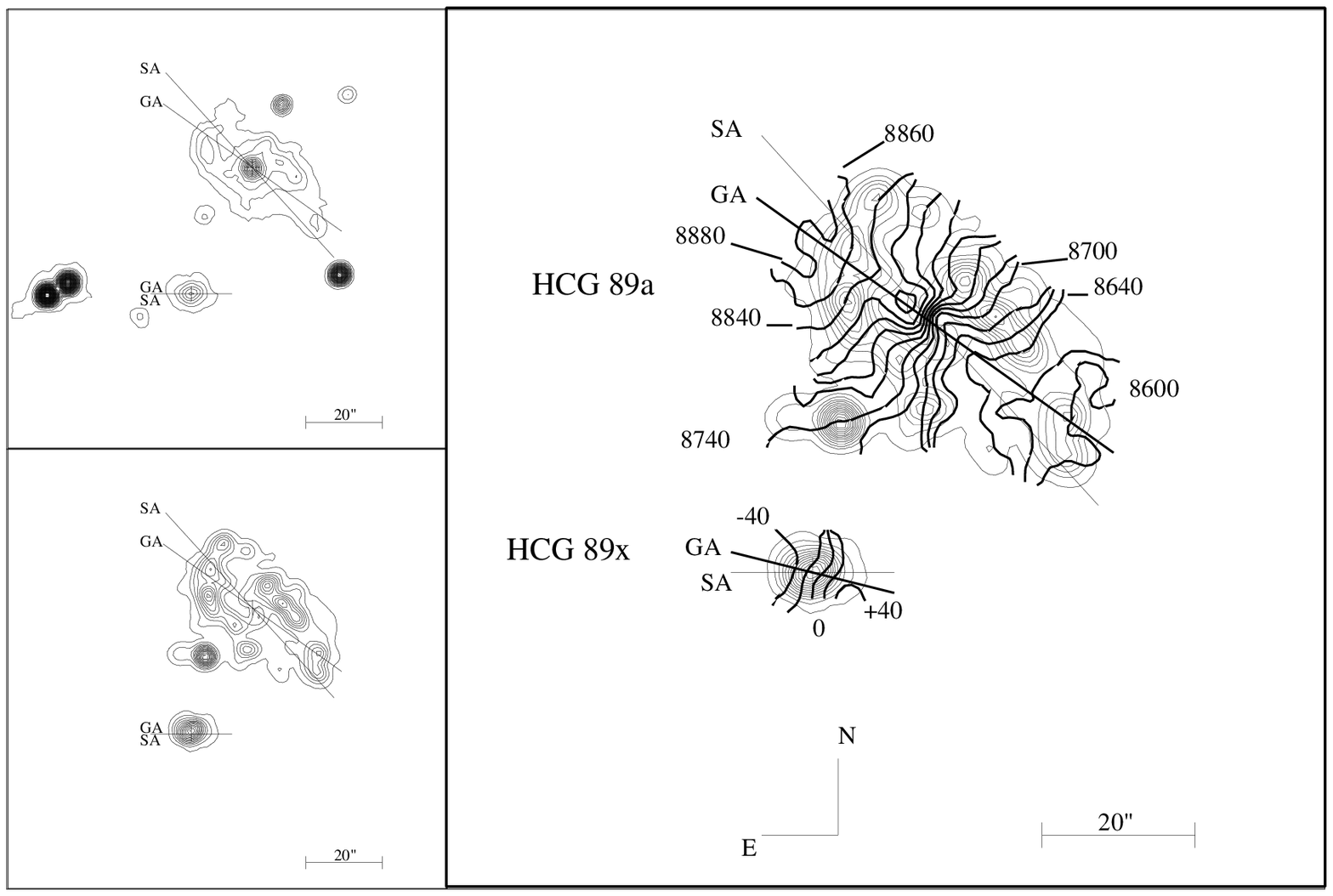}

\caption{HCG 89a: same as Fig. 1a.
Monochromatic and velocity maps for HCG 89x
are presented}

\figurenum{7b}

\epsfxsize=15cm \epsfbox{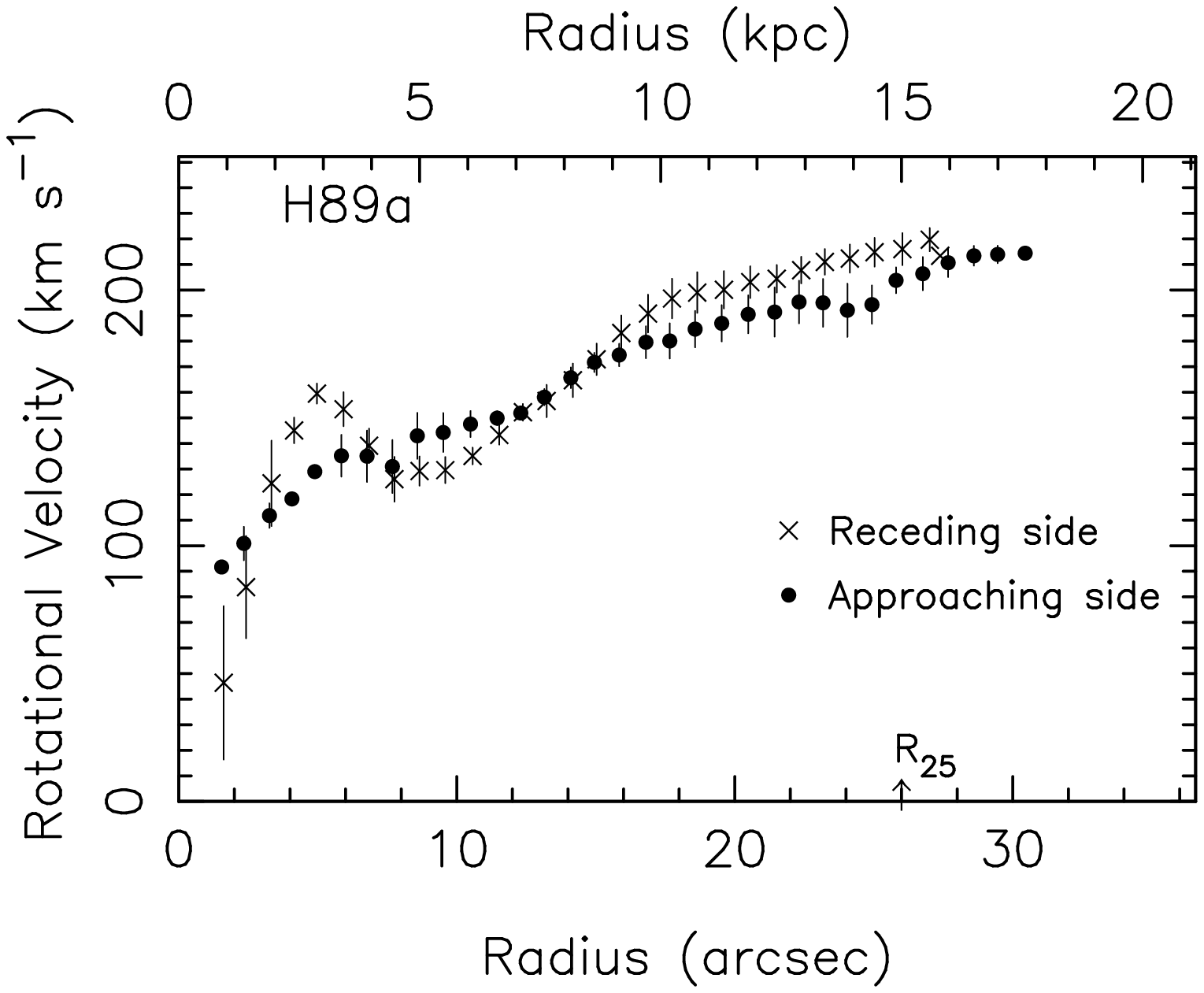}

\caption{HCG 89a: same as Fig.1b}

\end{figure*}

\clearpage


\begin{figure*}

\figurenum{8a} \vspace{-2.5cm}

\epsfxsize=17cm \epsfbox{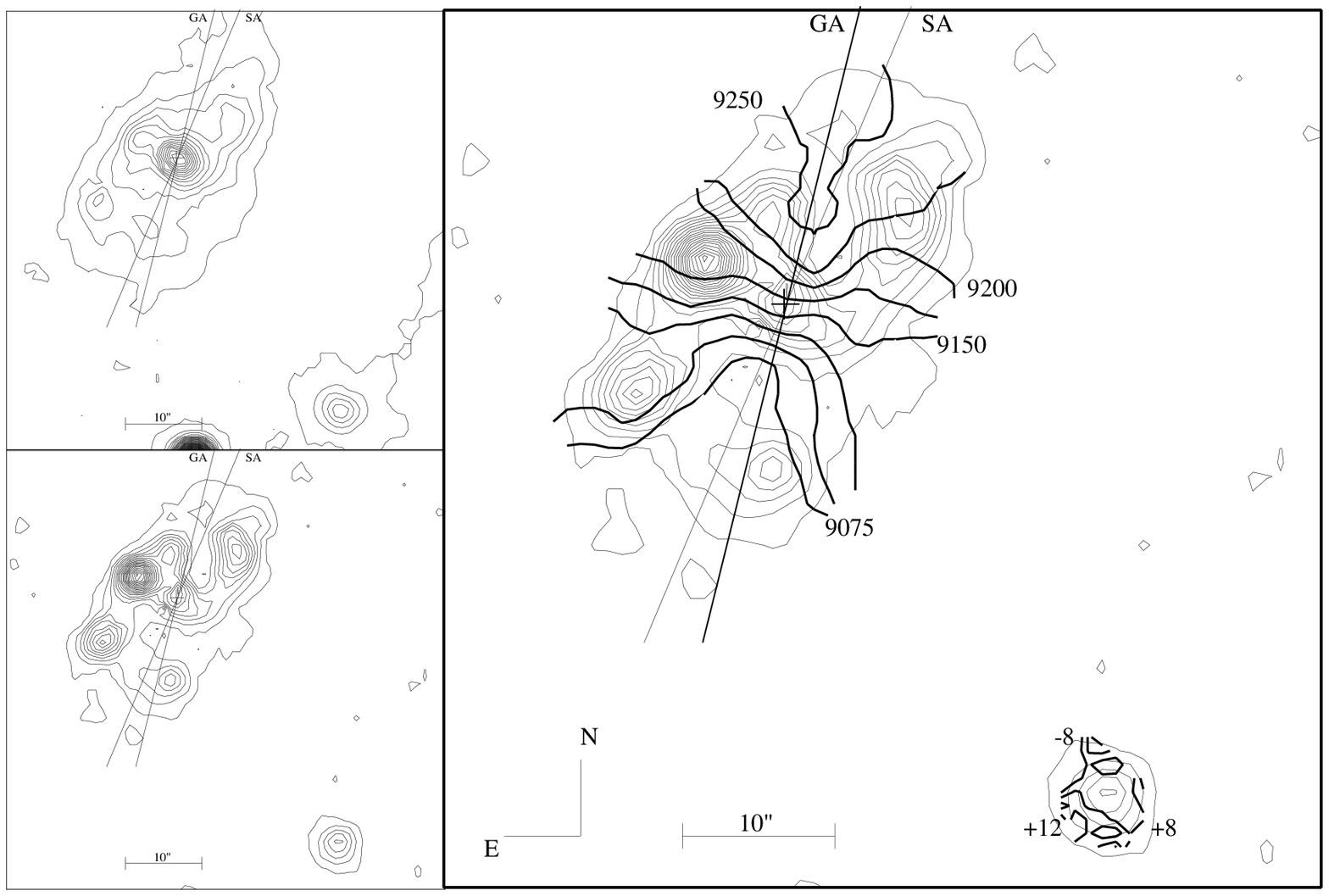}

\caption{ HCG 89b: same as Fig. 1a. Monochromatic and velocity maps for
an object on the south west of the field are presented.
}

\figurenum{8b}

\epsfxsize=15cm \epsfbox{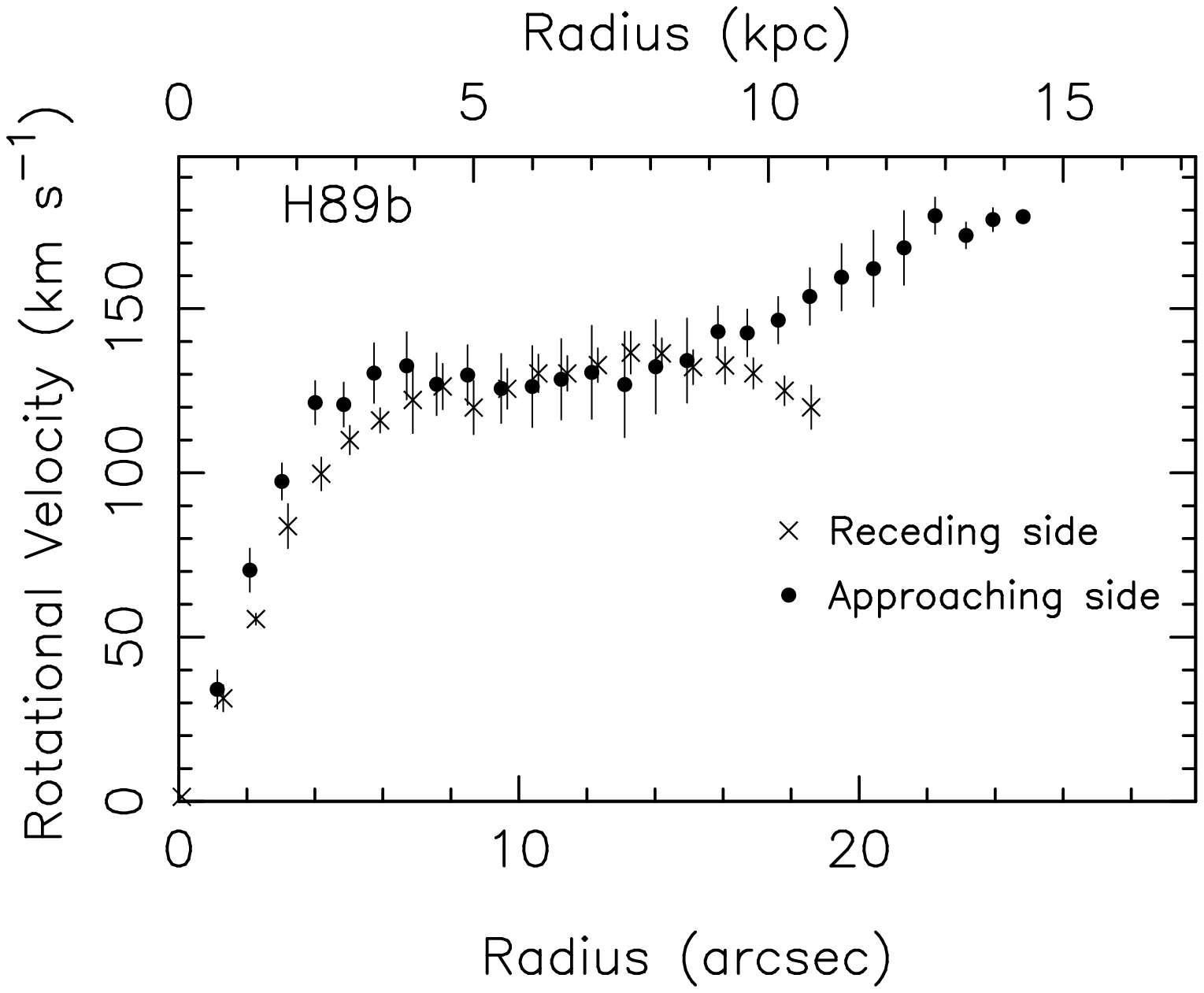}

\caption{HCG 89b: same as Fig.1b}

\end{figure*}

\clearpage


\begin{figure*}

\figurenum{9a} \vspace{-2.5cm}

\epsfxsize=15cm \epsfbox{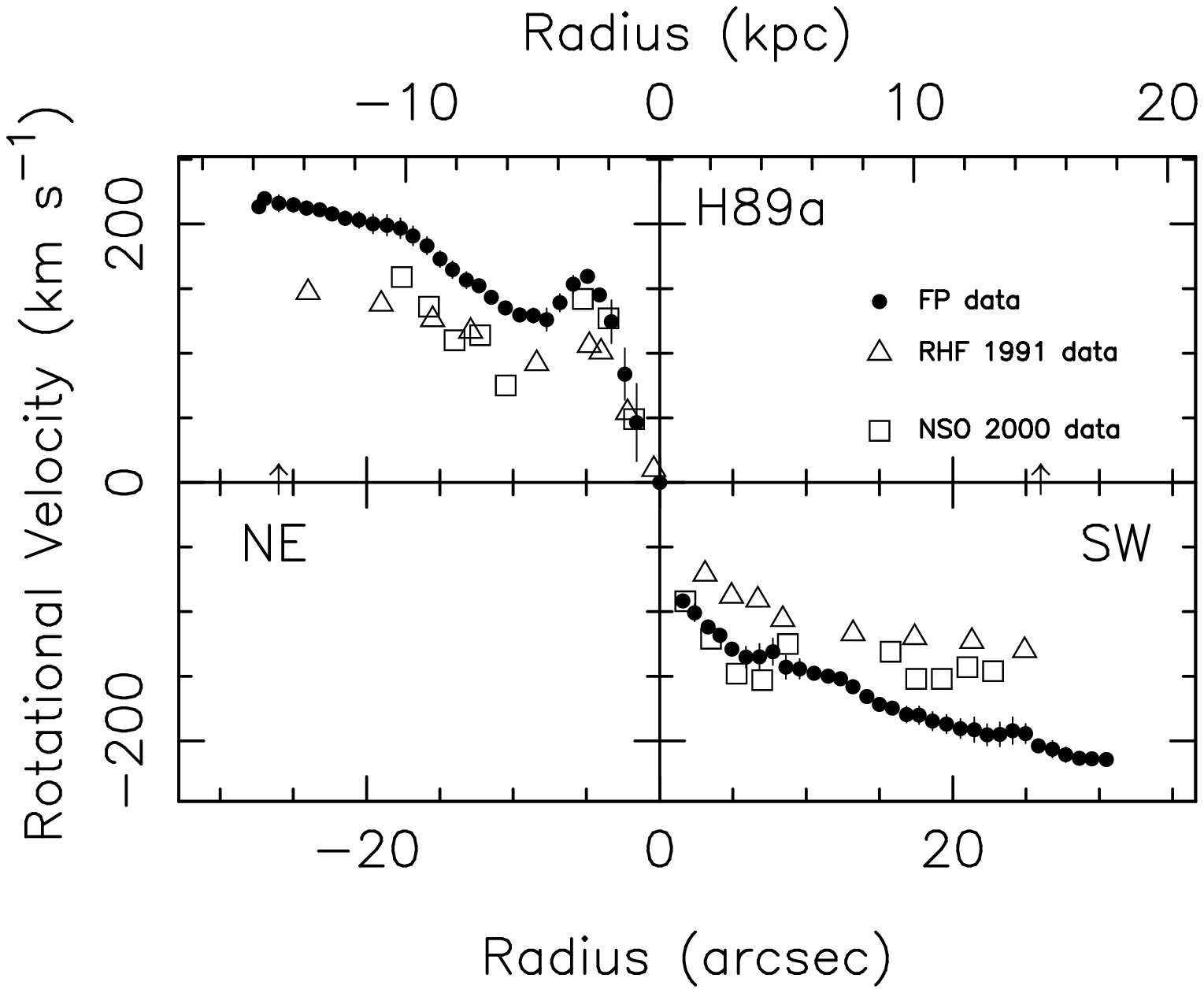}

\caption{HCG 89a: we are presenting different rotation curves using our
data, RHF1991 data and NSOMT2000 data. The kinematical parameters
used to plot the rotation curves are, for the inclination, $45^o$, $60^o$ and
$51^o$ respectively and for the position angle of the major axis,
$54^o$, $52^o$ and $57^o$ respectively.}

\figurenum{9b}

\epsfxsize=15cm \epsfbox{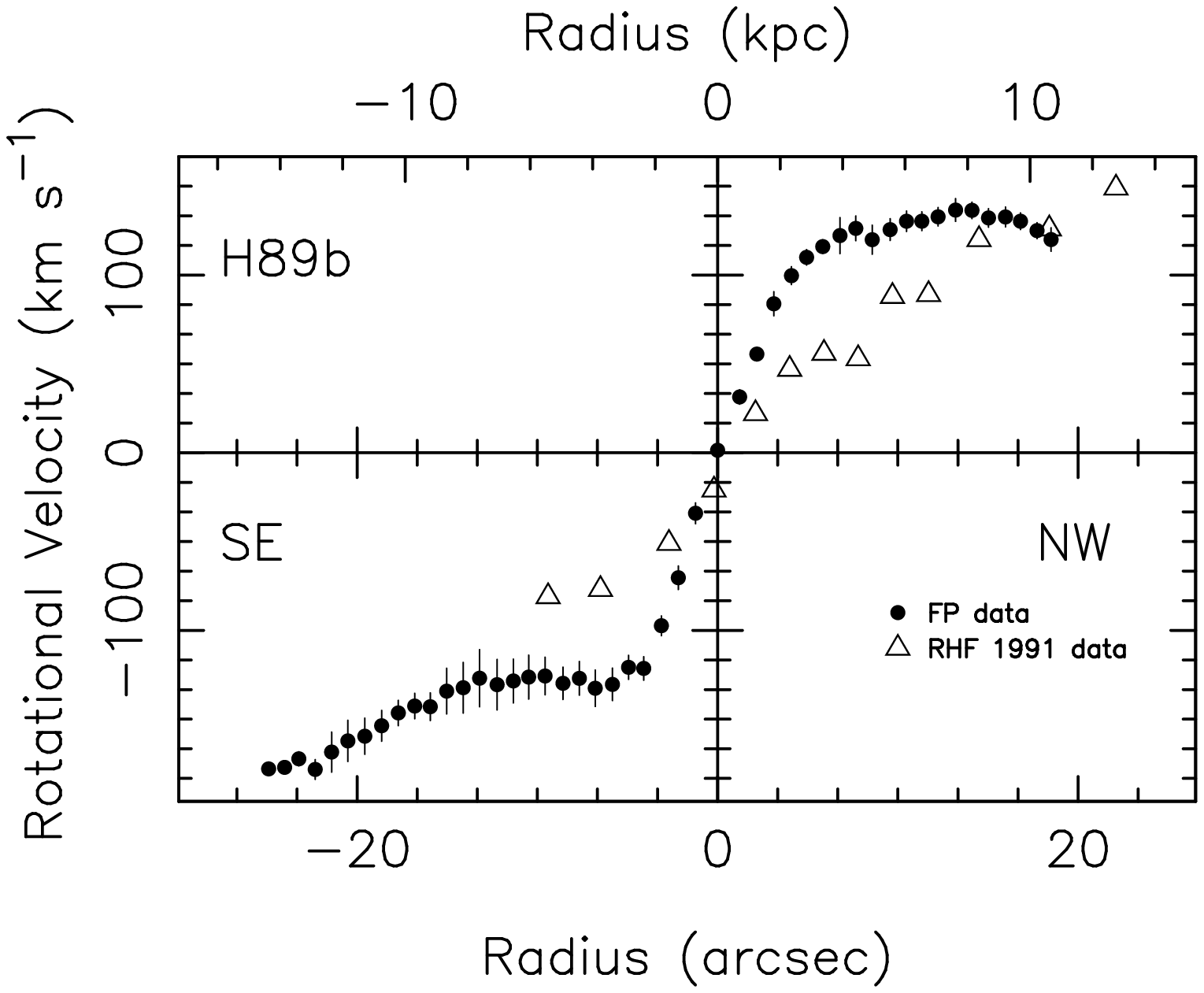}

\caption{HCG 89b: we are presenting different rotation curves using our
data and RHF1991 data. The kinematical parameters used to plot the
rotation curves are, for the inclination, $49^o$ and $63^o$ respectively and
for the position angle of the major axis, $165^o$ and $141^o$
respectively.}

\end{figure*}

\clearpage


\begin{figure*}

\figurenum{10a} \vspace{-2.5cm}

\epsfxsize=17cm \epsfbox{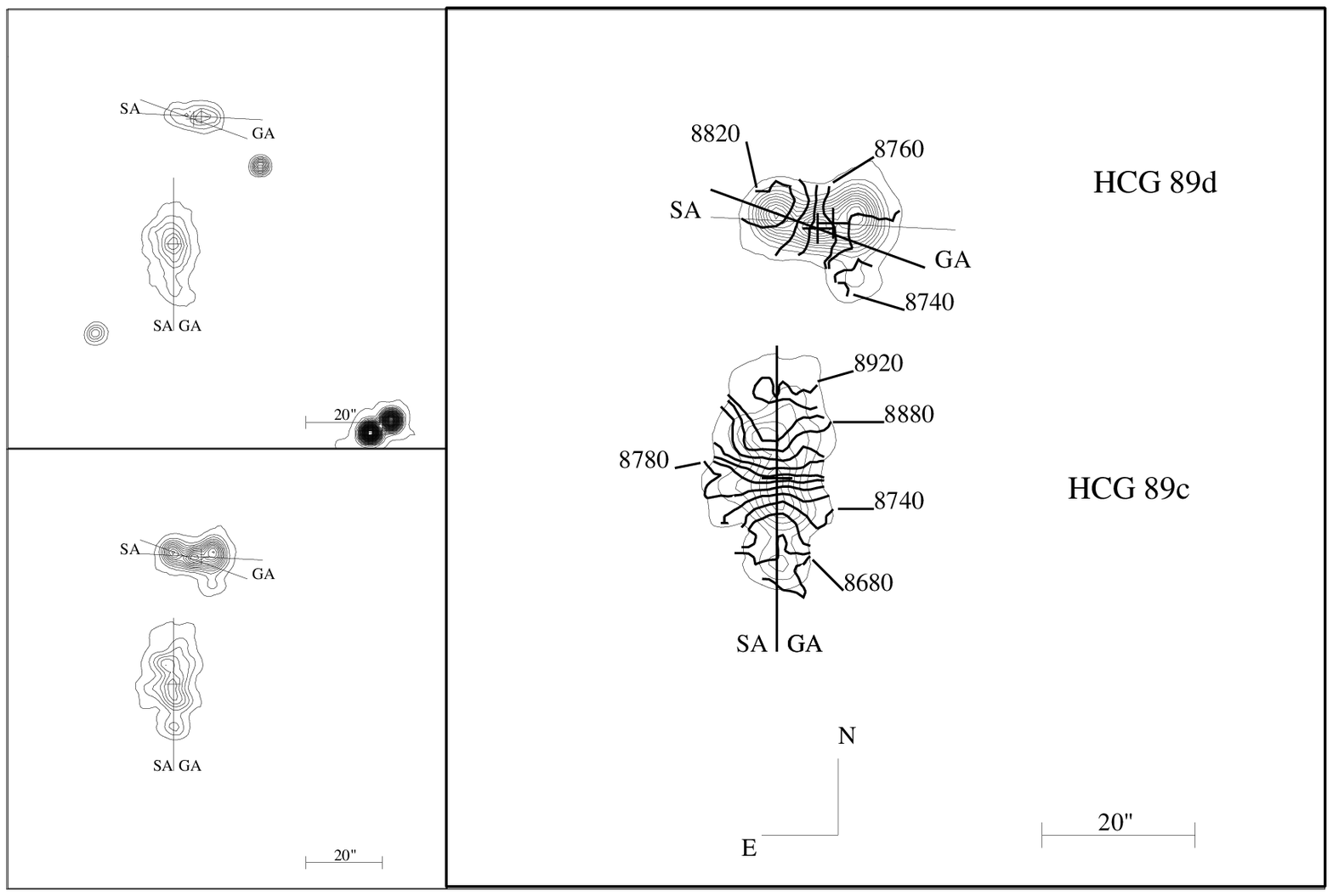}

\caption{HCG 89c-d: same as Fig.1a}

\figurenum{10b}

\epsfxsize=15cm \epsfbox{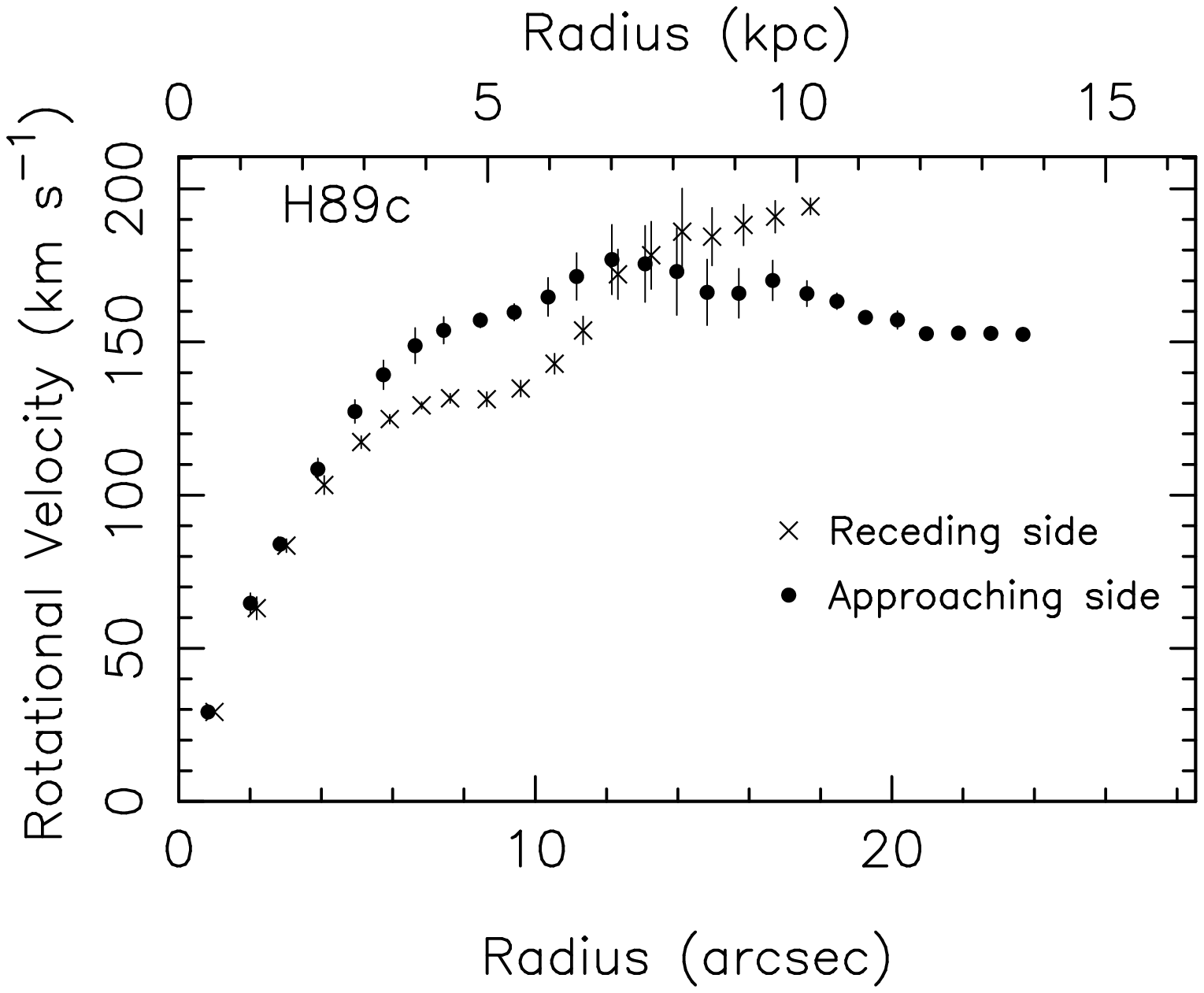}

\caption{HCG 89c: same as Fig.1b}

\end{figure*}

\clearpage


\begin{figure*}

\figurenum{11a} \vspace{-2.5cm}

\epsfxsize=17cm \epsfbox{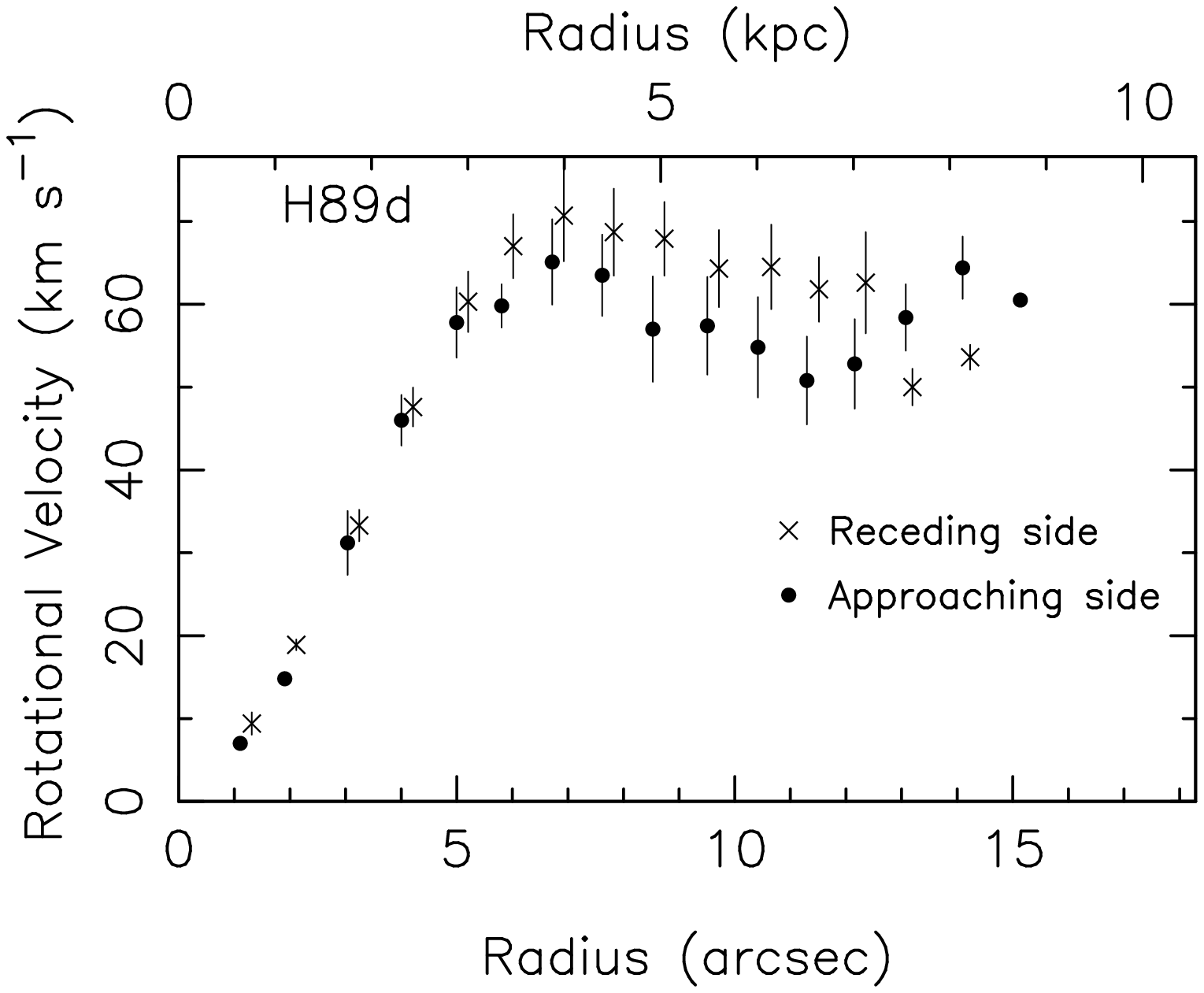}

\caption{HCG 89d: same as Fig.1b}

\figurenum{11b}

\epsfxsize=17cm \epsfbox{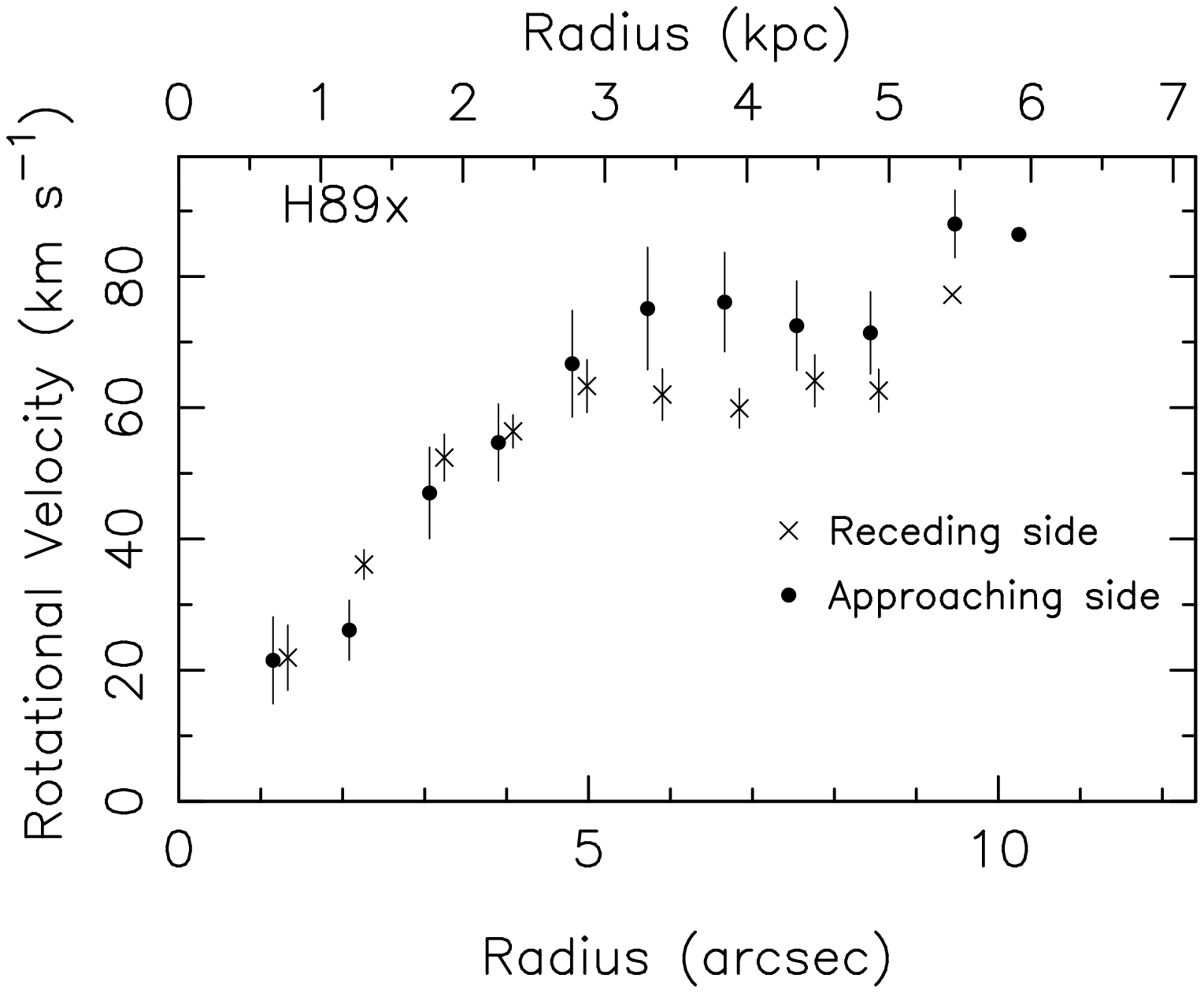}

\caption{HCG 89x: same as Fig.1b}

\end{figure*}

\clearpage



\begin{figure*}

\figurenum{12a} \vspace{-2.5cm}

\epsfxsize=17cm \epsfbox{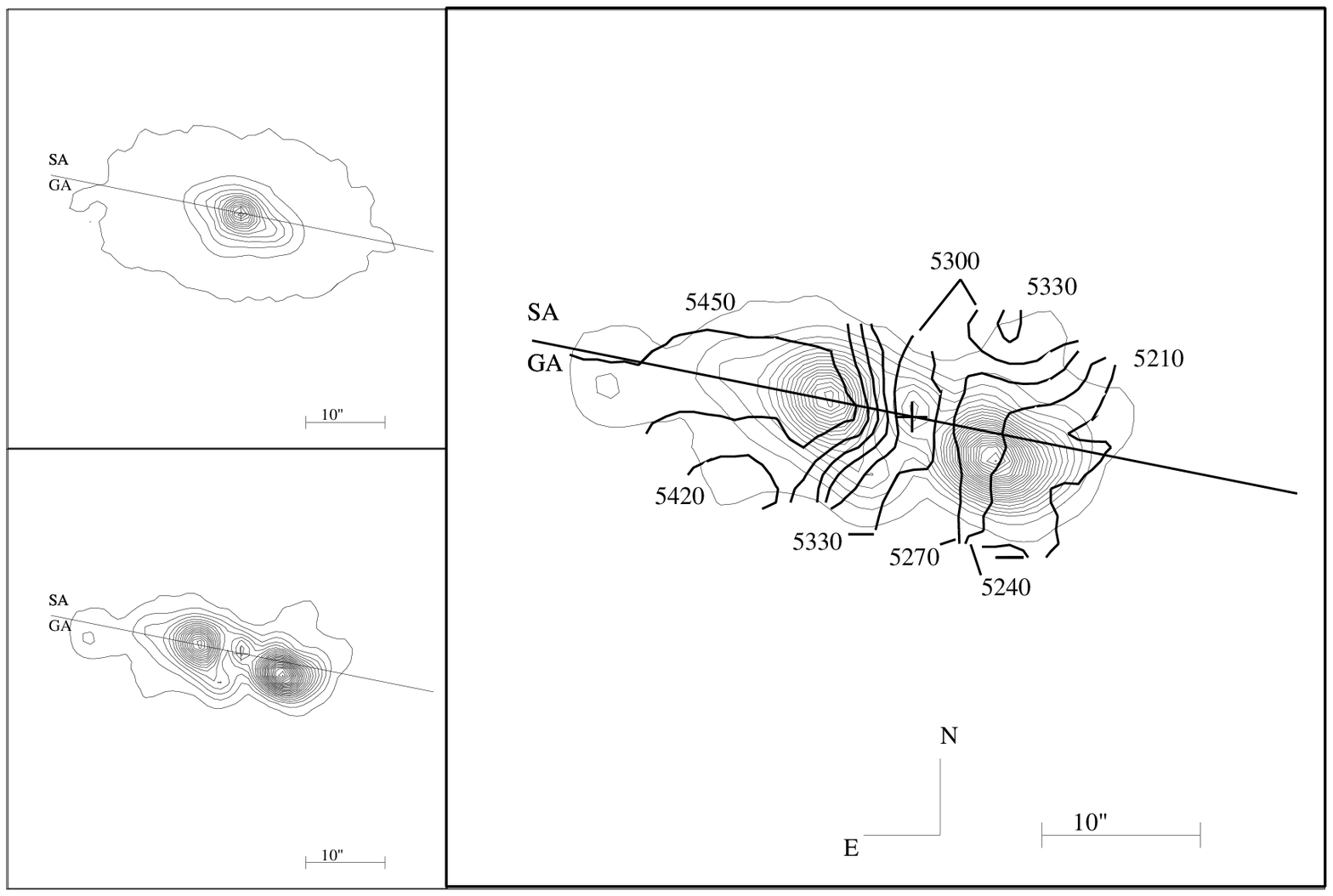}

\caption{HCG 100a: same as Fig.1a}

\figurenum{12b}

\epsfxsize=15cm \epsfbox{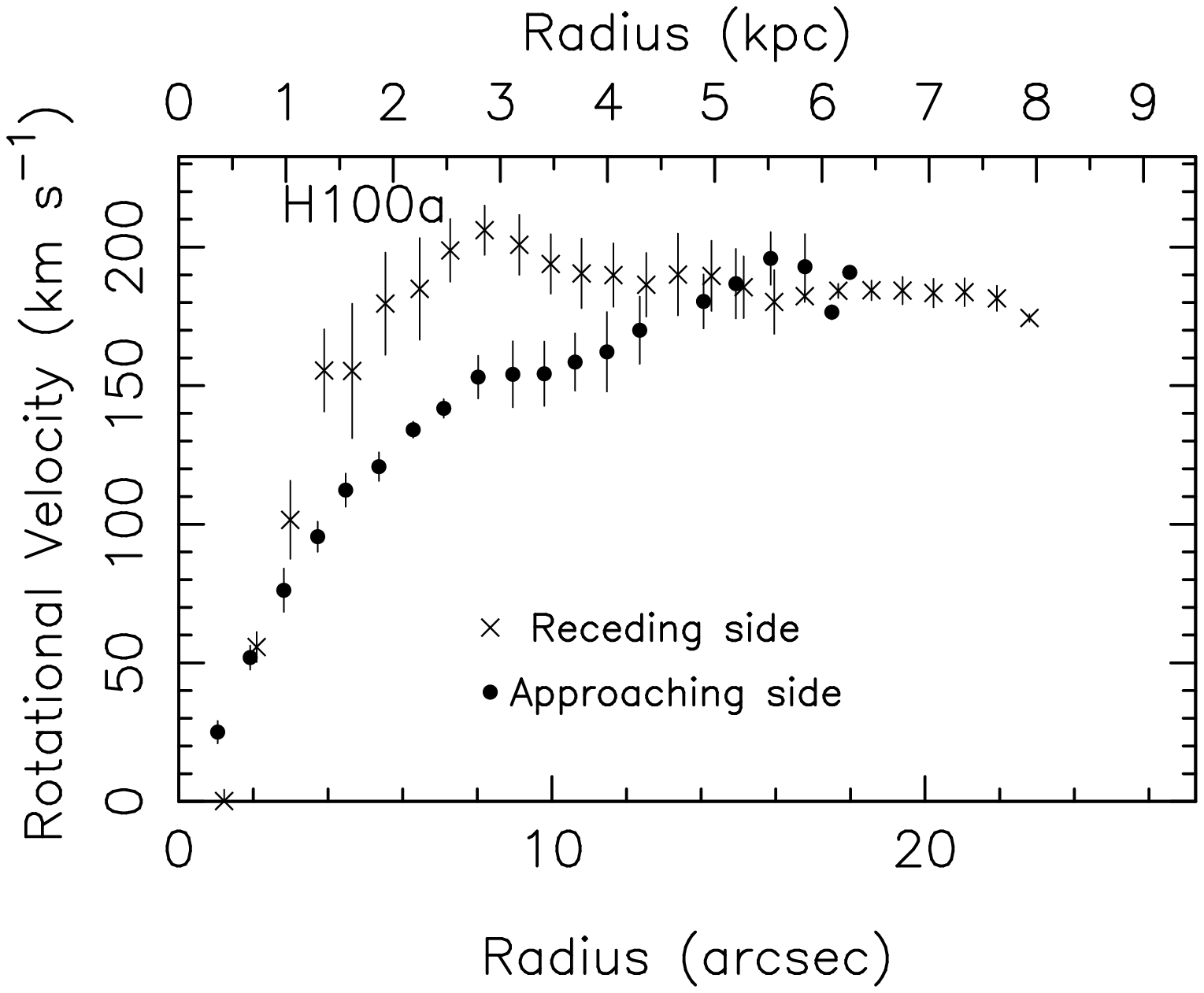}

\caption{HCG 100a: same as Fig.1b}

\end{figure*}

\clearpage


\begin{figure*}

\figurenum{13a} \vspace{-2.5cm}

\epsfxsize=17cm \epsfbox{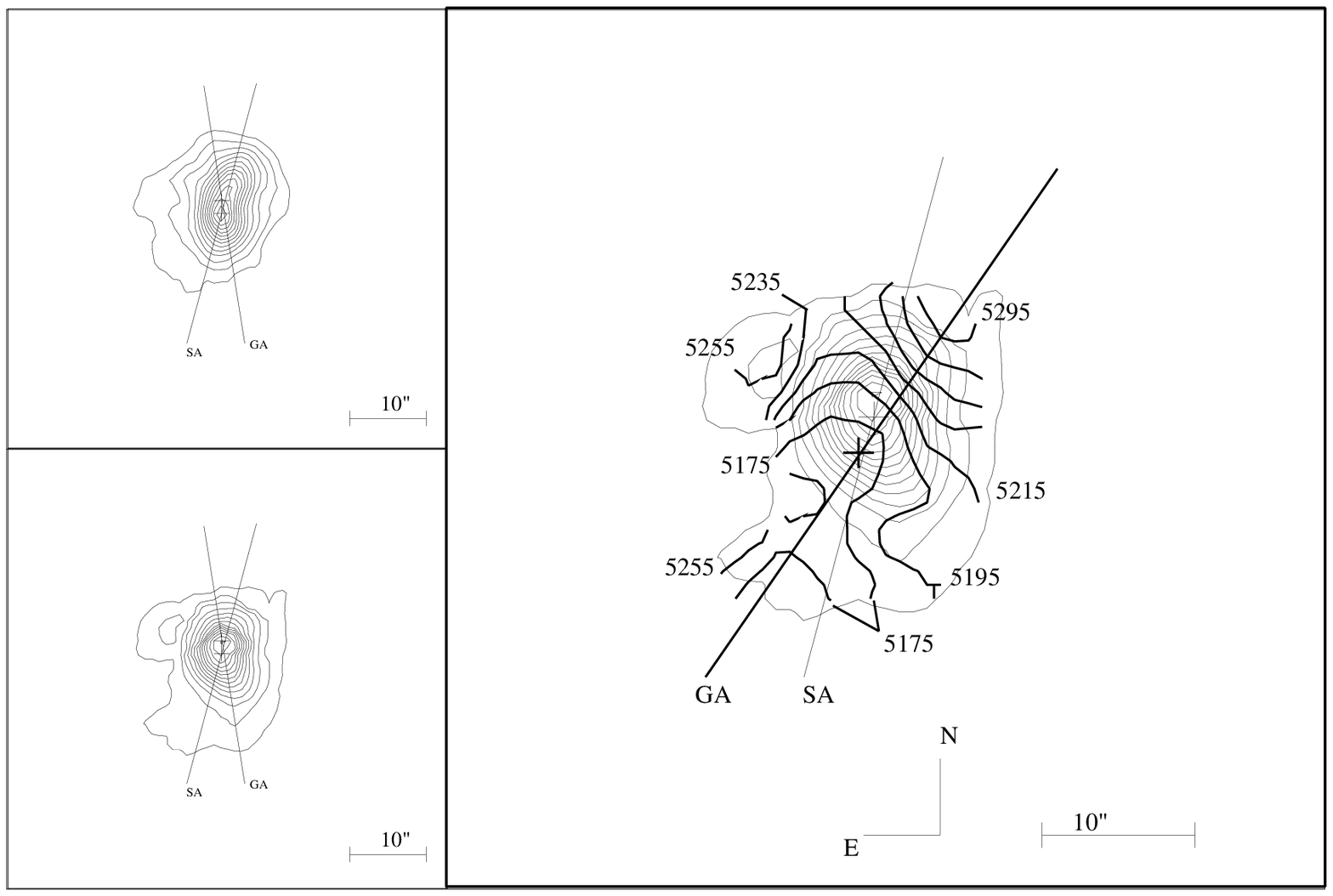}

\caption{HCG 100b: same as Fig.1a}

\figurenum{13b}

\epsfxsize=15cm \epsfbox{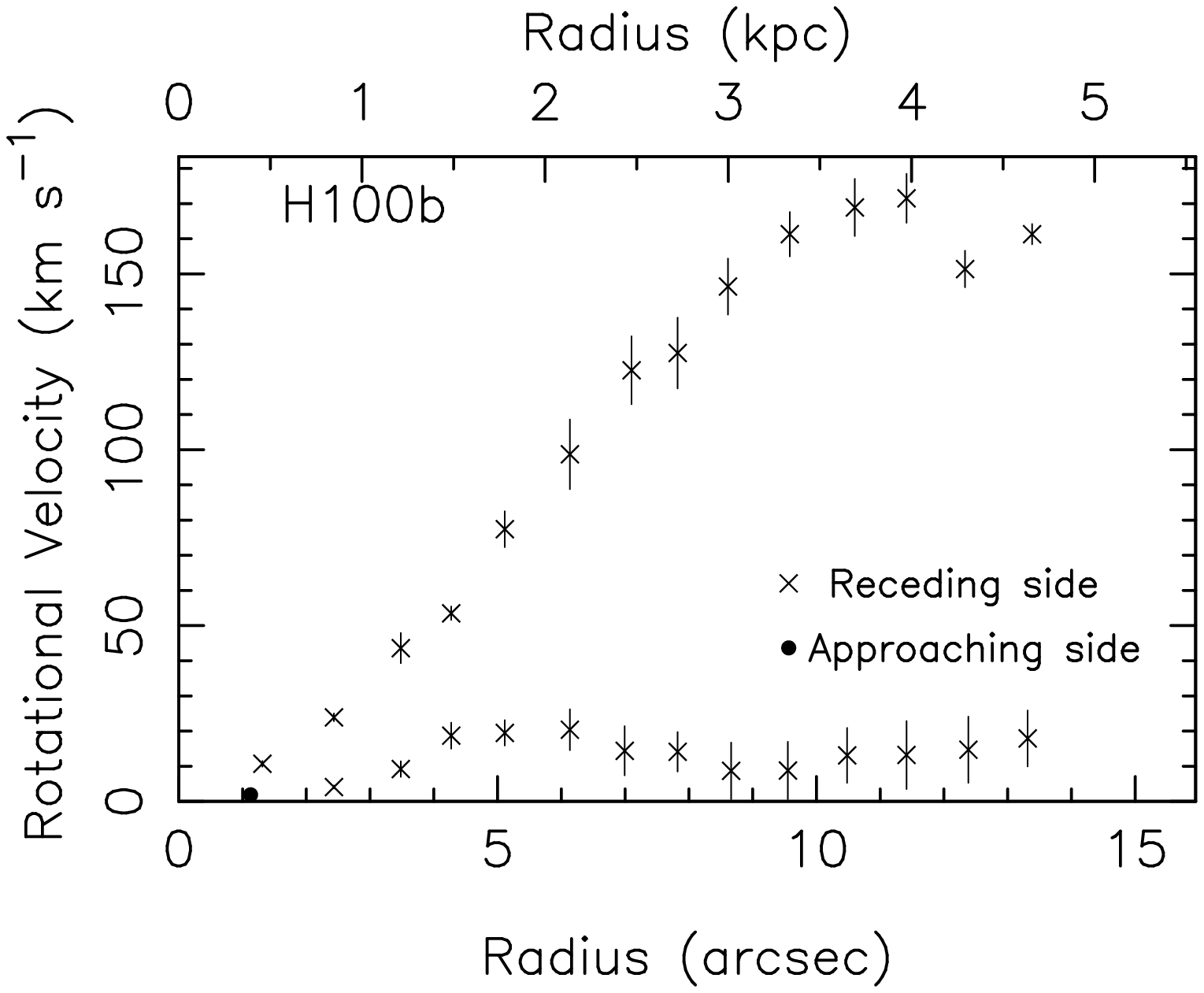}

\caption{HCG 100b: same as Fig.1b}

\end{figure*}

\clearpage


\begin{figure*}

\figurenum{14a} \vspace{-2.5cm}

\epsfxsize=17cm \epsfbox{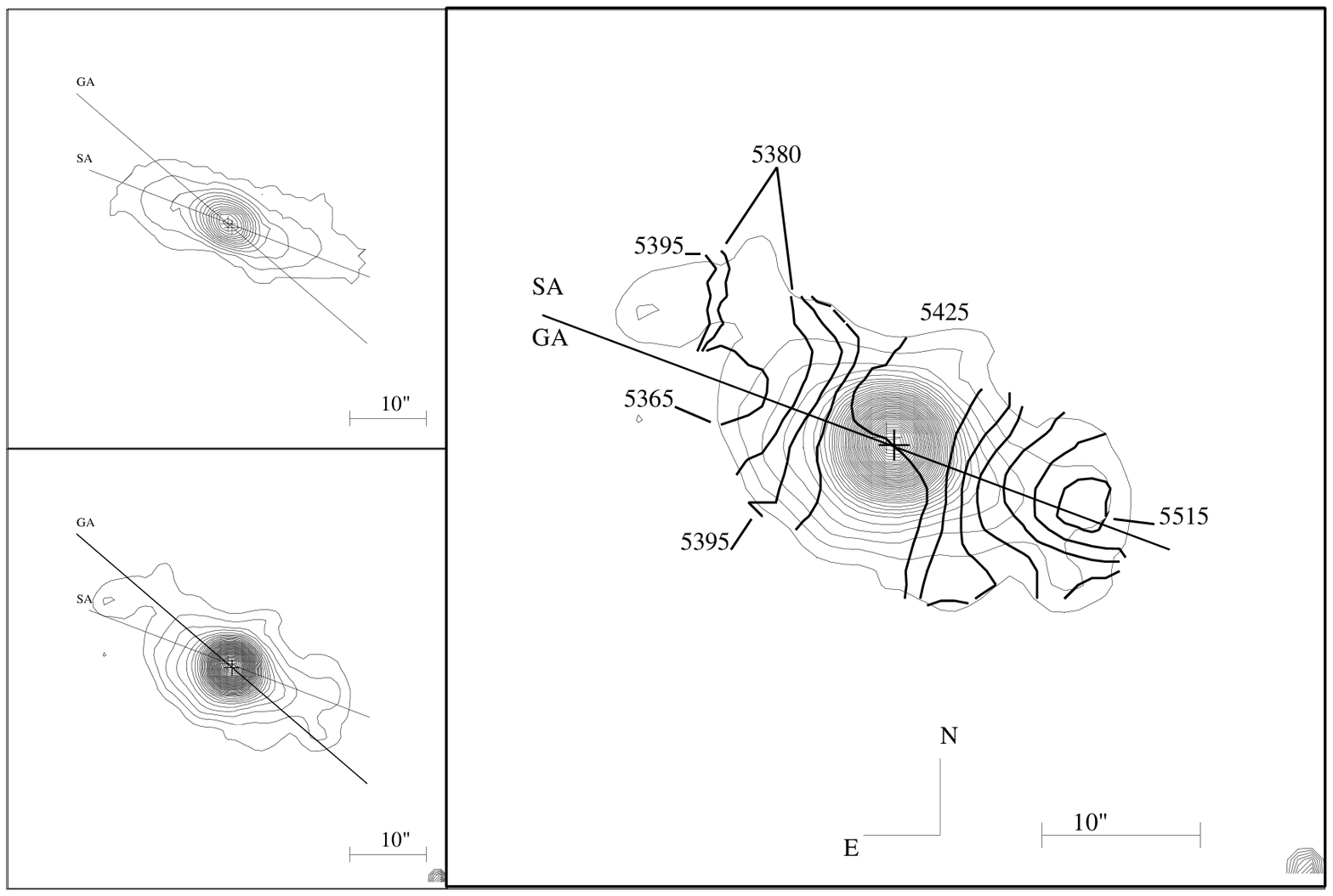}

\caption{HCG 100c: same as Fig.1a}

\figurenum{14b}

\epsfxsize=15cm \epsfbox{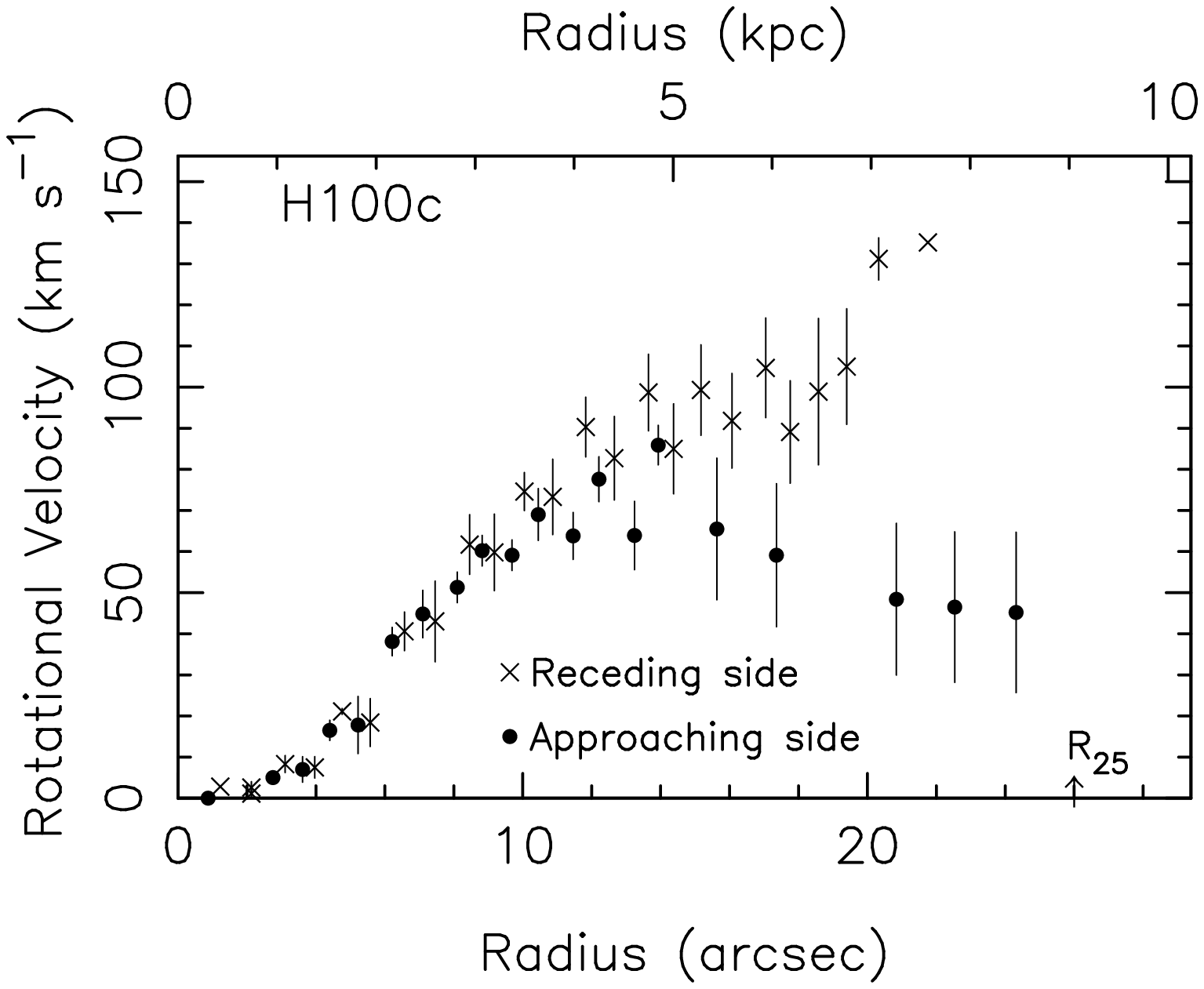}

\caption{HCG 100c: same as Fig.1b}

\end{figure*}

\clearpage


\begin{figure*}

\figurenum{15a} \vspace{-2.5cm}

\epsfxsize=17cm \epsfbox{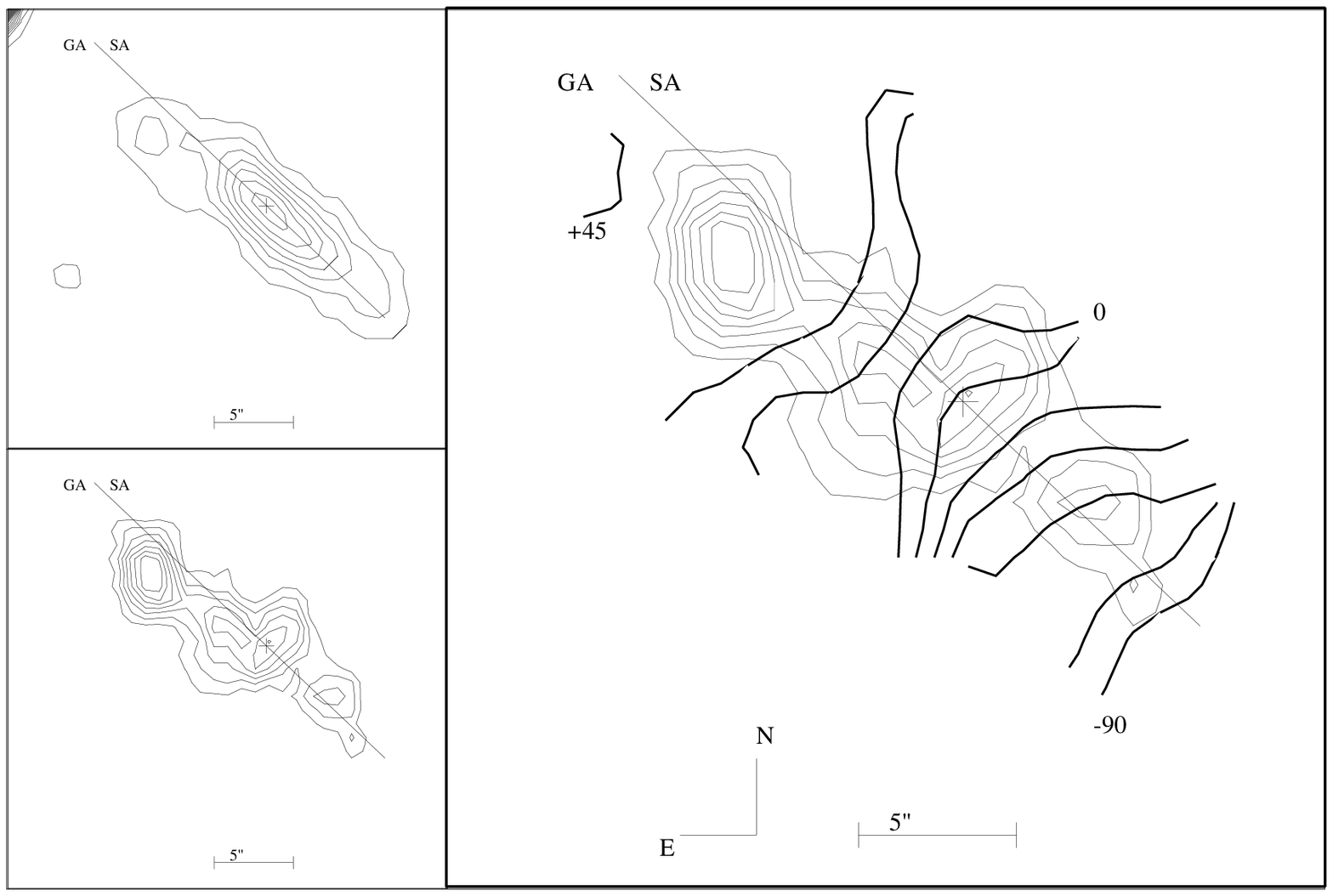}

\caption{HCG 100d: same as Fig.1a}

\figurenum{15b}

\epsfxsize=15cm \epsfbox{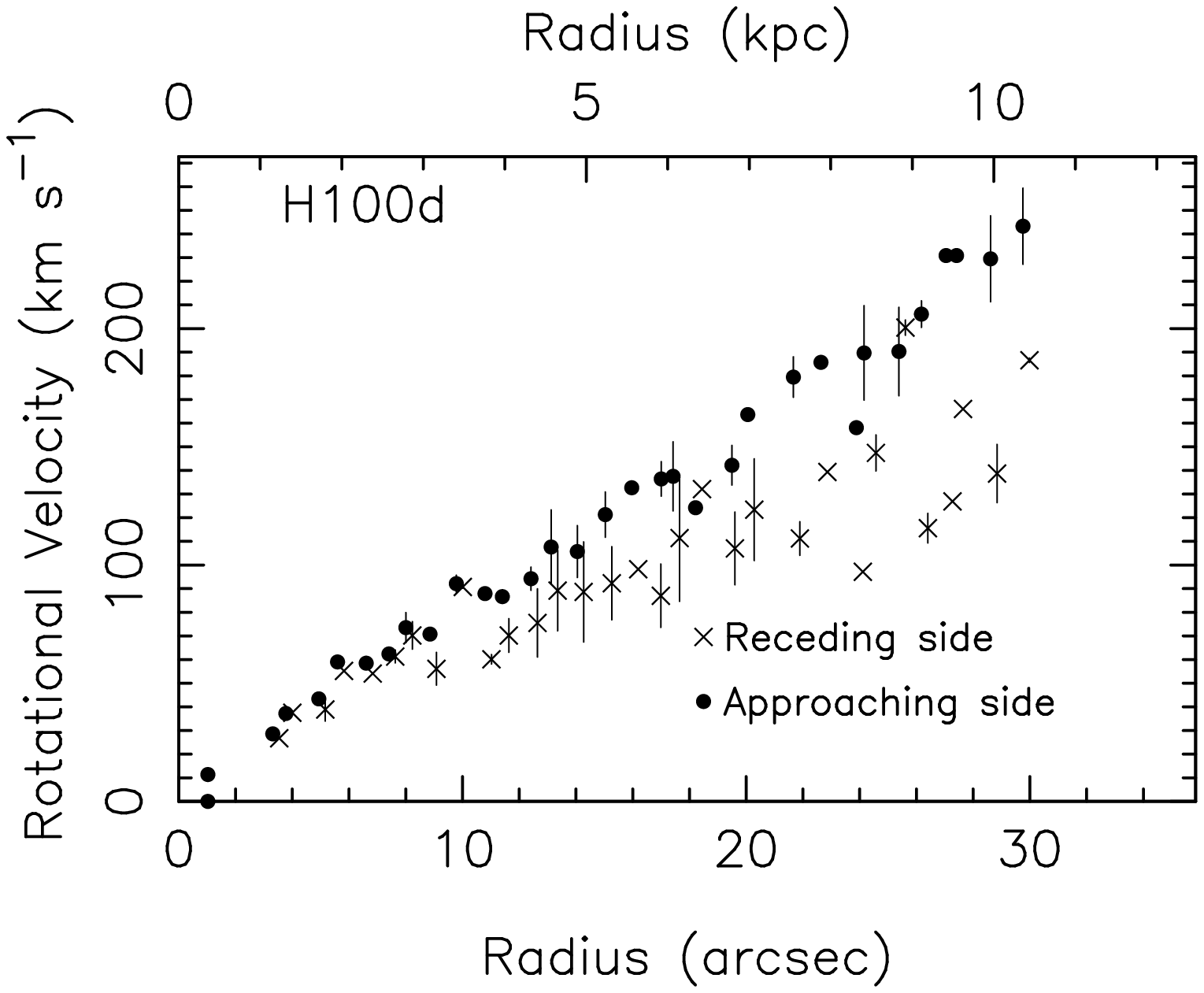}

\caption{HCG 100d: same as Fig.1b}

\end{figure*}

\clearpage


\begin{figure*}

\figurenum{16a} \vspace{-2.5cm}

\epsfxsize=15cm \epsfbox{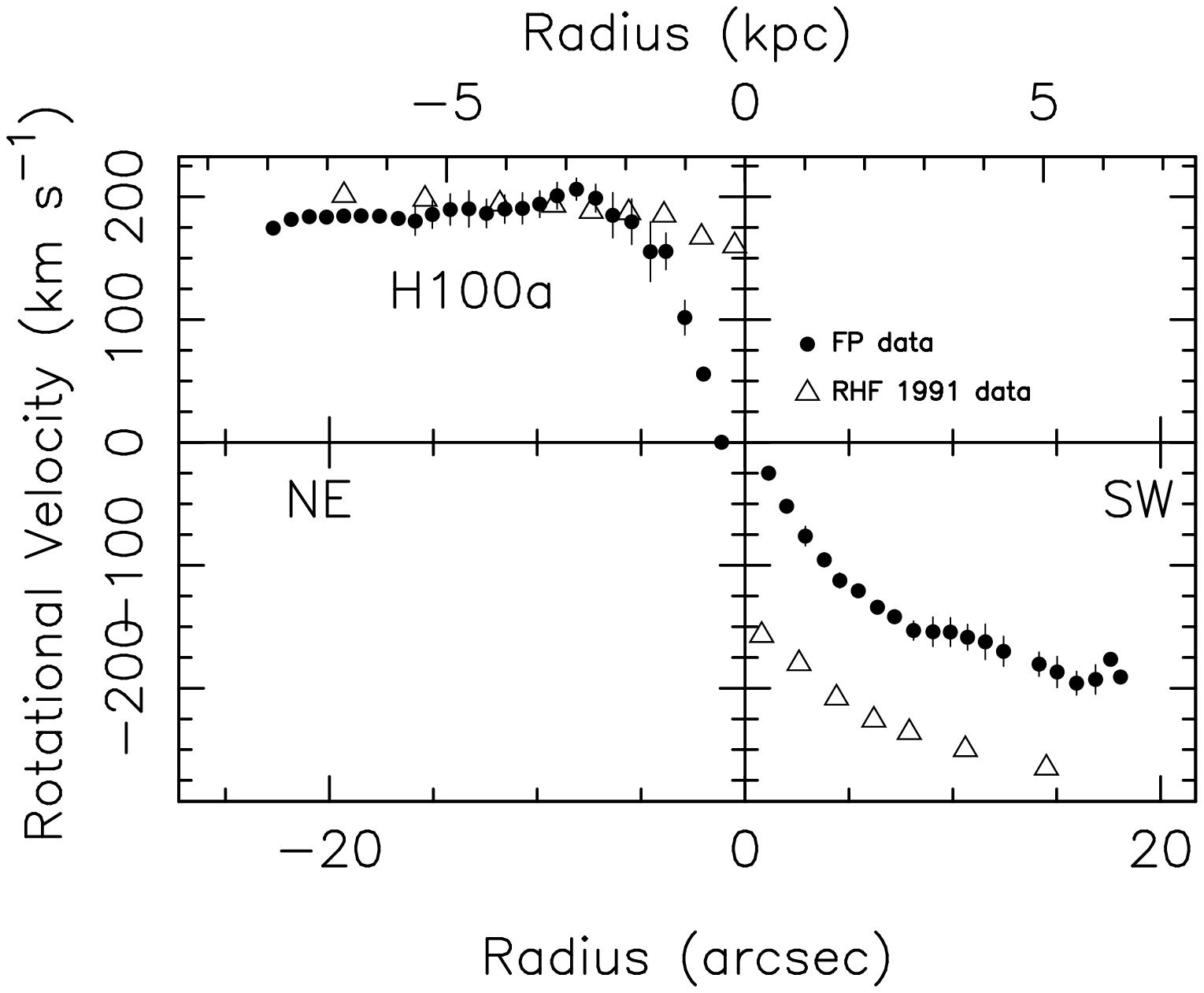}

\caption{HCG 100a: we are presenting different rotation curves
using our data and RFH1991 data. The kinematical parameters used
to plot the rotation curves are, for the inclination, $50^o$ and $60^o$
respectively and for the position angle of the major axis, $78^o$
and $85^o$ respectively.}

\figurenum{16b}

\epsfxsize=15cm \epsfbox{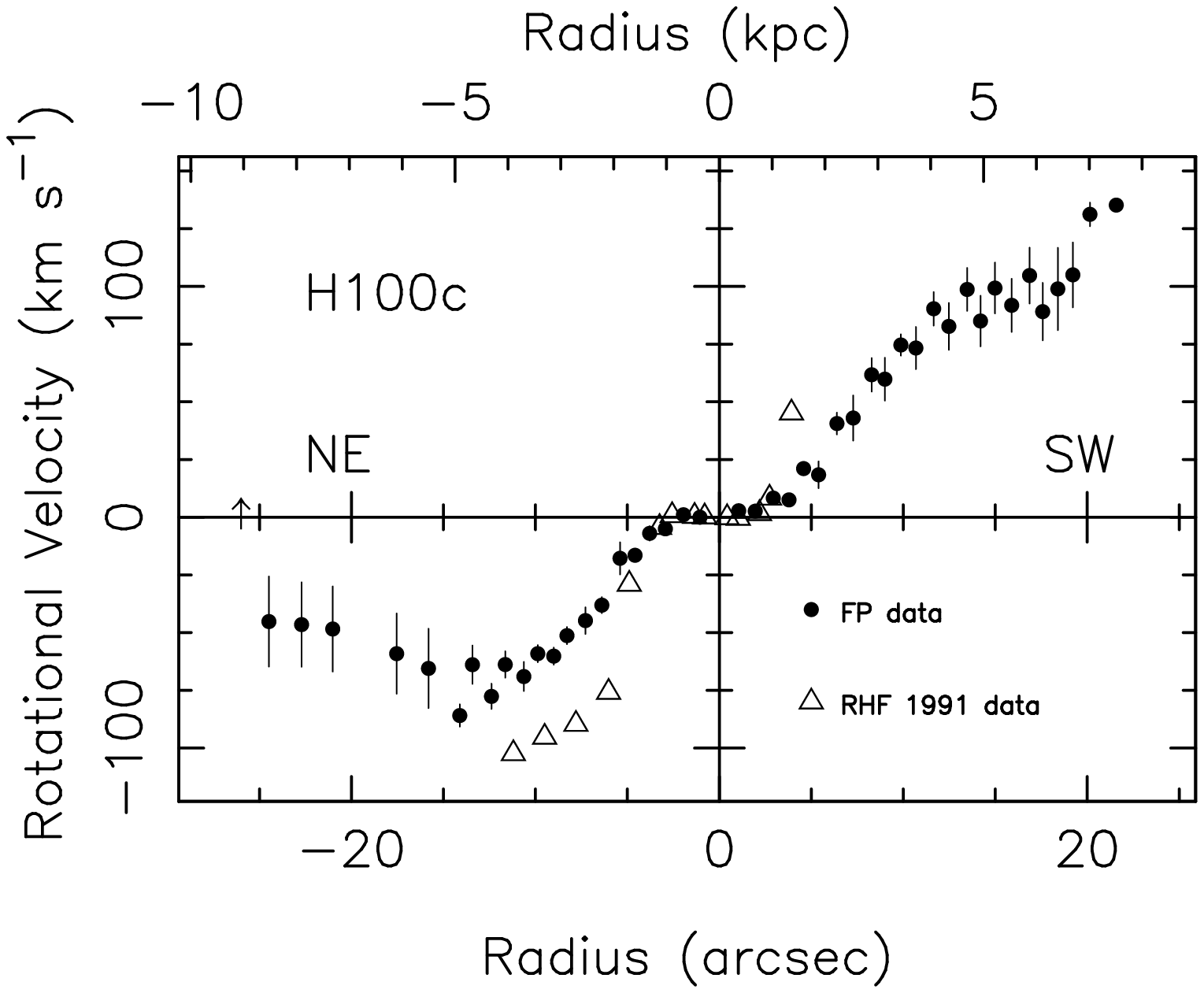}

\caption{HCG 100c: we are presenting different rotation curves
using our data and RFH1991 data.  The kinematical parameters used
to plot the rotation curves are, for the inclination, $66^o$ and $68^o$
respectively and for the position angle of the major axis, $72^o$
and $73^o$ respectively.}

\end{figure*}

\clearpage


\begin{figure*}

\figurenum{17a} \vspace{-2.5cm}

\epsfxsize=17cm \epsfbox{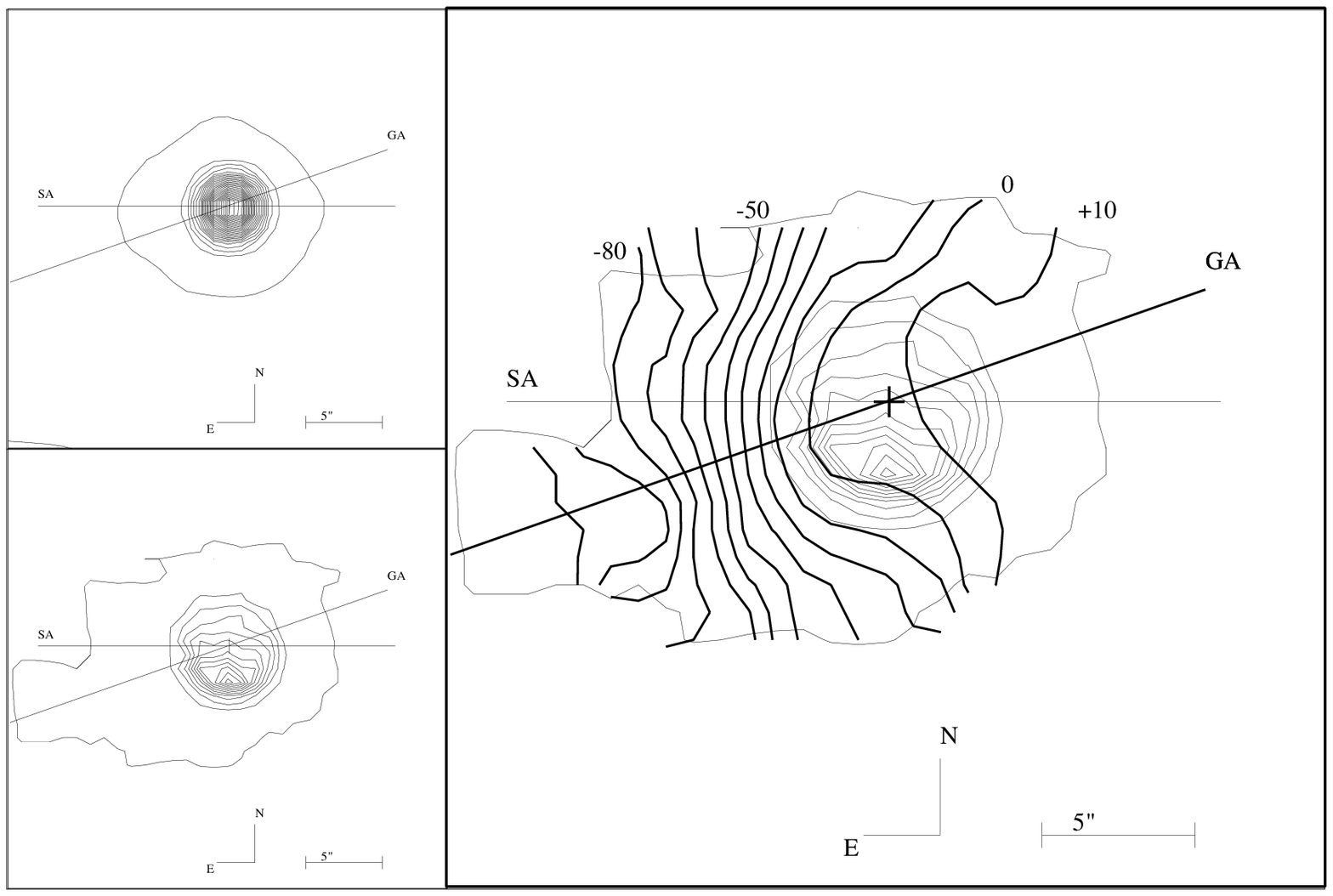}

\caption{HCG 100x: same as Fig.1a}

\figurenum{17b}

\epsfxsize=15cm \epsfbox{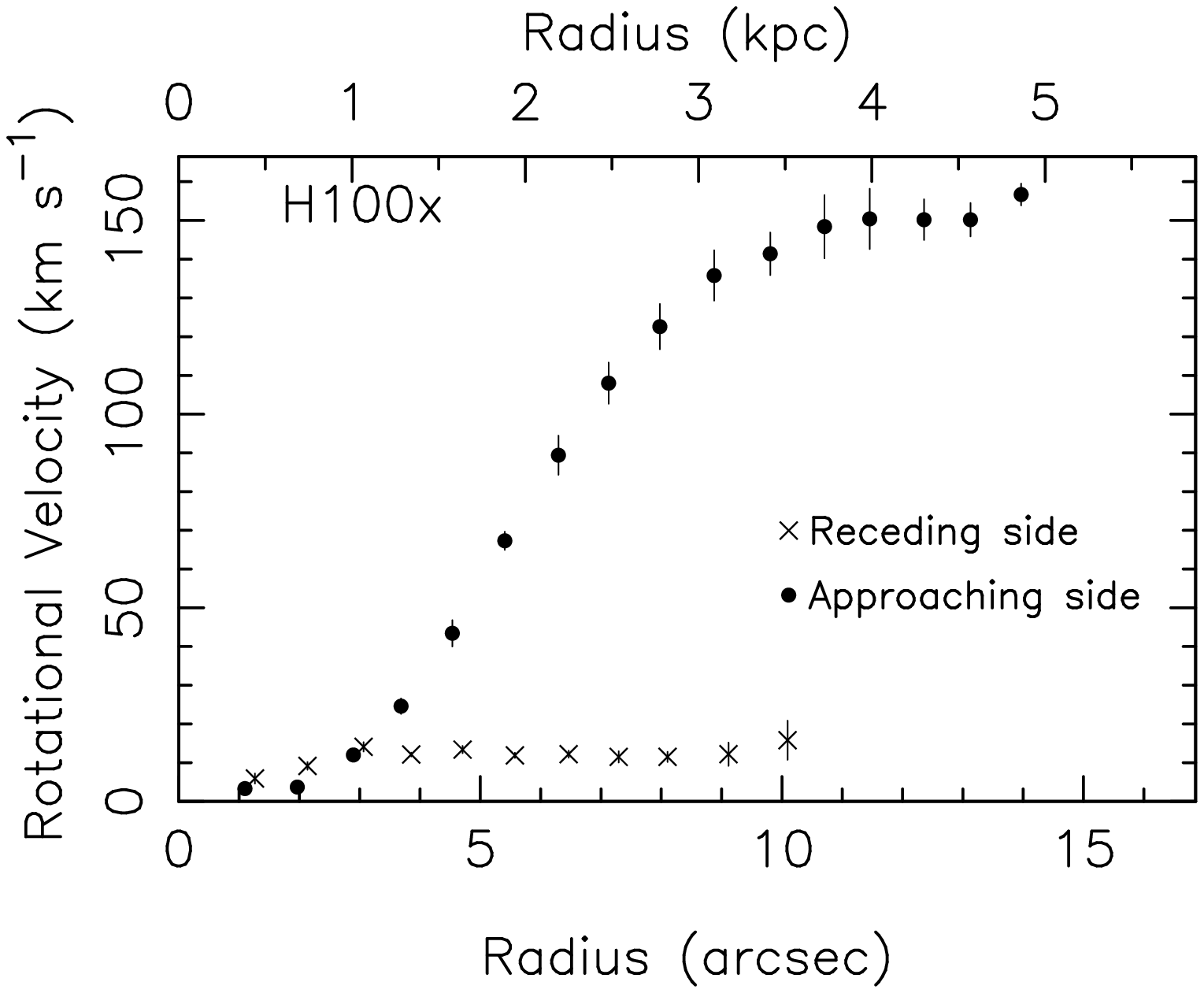}

\caption{HCG 100x: same as Fig.1b}

\end{figure*}

\clearpage

%
%

\begin{deluxetable}{llll}


\small \tablenum{1} \tablecolumns{4} \tablewidth{0pt}
\tablecaption{Journal of Perot-Fabry observations \label{tbl-1}}

\tablehead{ \colhead{} & \colhead{Hickson Compact Groups} &
\colhead{}}
\startdata

Observations & Telescope & CFHT 3.6m & ESO 3.6m \\

             & Equipment & MOS/FP @ Cassegrain & CIGALE @ Cassegrain\\

             & Date & 1996, August, 27-29 & 1995, August, 21-24 \\

             & Seeing & $<$ 1" & $\sim$ 1" \\

Calibration & Neon Comparison light & $\lambda$ 6598.95 \AA &
 $\lambda$ 6598.95 \AA \\

Perot--Fabry & Interference Order & 1162 @ 6562.78 \AA & 796 @ 6562.78 \AA \\

         & Free Spectral Range at H$\alpha$ & $265~km~s^{-1}$ & $380~km~s^{-1}$ \\

         & Finesse at H$\alpha$ & 12 & 12\\

         & Spectral resolution at H$\alpha$ & 13672 \tablenotemark{1} & 9375 \tablenotemark{1}\\

Sampling & Number of Scanning Steps & 24 & 24 \\

     & Sampling Step & 0.24 \AA\ ($11~km~s^{-1}$) & 0.35 \AA\ ($16~km~s^{-1}$)\\

     & Total Field & 440''$\times $440'' (512$\times $512 px$^2$)\tablenotemark{2} & 170''$\times $170'' (256$\times $256 px$^2$) \\

         & Pixel Size (binned) & 0.86'' & 0.91''  \\

Detector & & STIS 2 CCD & IPCS  \\

\enddata

\tablenotetext{1}{For a signal to noise ratio of 5 at the sample
step }

\tablenotetext{2}{After binning 2*2}

\end{deluxetable}

\clearpage

%
%

\begin{deluxetable}{lcccccc}
\footnotesize

\tablenum{2} \tablewidth{0pt}

\tablecaption{Observational Characteristics}

\tablecolumns{7}

\tablehead{
\colhead{Name} & \colhead{Telescope} & \multicolumn{2}{c}{Exposure Time} & \multicolumn{3}{c}{Interference Filter}  \\

\colhead{} & \colhead{} & \colhead{} & \colhead{} & \colhead{} & \colhead{} & \colhead{} \\

\colhead{} & \colhead{} & \colhead{Total} & \colhead{per channel} & \colhead{Central wavelength} & \colhead{FWHM} & \colhead{Transmission} \\
\colhead{} & \colhead{} & \colhead{} & \colhead{} & \colhead{} &
\colhead{} & \colhead{at maximum}}

\startdata

HCG 88 ab& ESO 3.6m & 2h & 300s & 6688 \AA & 22 \AA & 0.73 \\

\hspace{1.4cm}cd& CFHT 3.6m & 1.2h & 180s & 6697 \AA & 17 \AA & 0.73 \\

         &  &  &  &  &  &  \\

HCG 89   & ESO 3.6m & 2h & 300s & 6763 \AA & 30 \AA & 0.73 \\

         &  &  &  &  &  &  \\

HCG 100acd& CFHT 3.6m & 1.2h & 180s & 6697 \AA & 17 \AA & 0.73 \\

\hspace{1.4cm}b & ESO 3.6m & 2h & 300s & 6668 \AA & 20 \AA & 0.60 \\

\enddata

\end{deluxetable}

\clearpage

%
%

\begin{deluxetable}{rcccccc}
\small

\tablenum{3}

\tablecolumns{6}

\tablewidth{0pt}

\tablecaption{Properties of HCGs members}

\tablehead{ \colhead{Name} & \colhead{$\alpha$ (1950)} &

\colhead{$\delta$ (1950)}& \colhead{Morphological\tablenotemark{1}
}&

\colhead{Systemic\tablenotemark{1} } & \colhead{B$_{TC}$\tablenotemark{1} } & \colhead{R$_{25}$\tablenotemark{2} } \\

\colhead{} & \colhead{} & \colhead{}& \colhead{Type} &
\colhead{Velocity ($km~s^{-1}$)} & \colhead{} & \colhead{(Arcsec)}}

\startdata

HCG 88 a & 20h 49m 55.6s & -05\Deg 53\Min 59.7\Sec & Sb & 6033 & 13.18 & 44  \\

       b & 20h 49m 51.0s & -05\Deg 56\Min 08.5\Sec & SBb & 6010 & 13.24 & 37 \\

       c & 20h 49m 47.2s & -05\Deg 57\Min 40.8\Sec & Sc & 6083 & 13.87 & 38  \\

       d & 20h 49m 33.9s & -05\Deg 59\Min 12.7\Sec & Sc & 6032 & 14.49 & 33  \\

         &  &  &  &  &  &   \\

HCG 89 a & 21h 17m 24.3s & -04\Deg 08\Min 04.4\Sec & Sc & 8850  &
14.10 & 26   \\
       b & 21h 17m 42.5s & -04\Deg 06\Min 31.8\Sec & SBc & 8985 & 14.88 & - \\

       c & 21h 17m 31.6s & -04\Deg 07\Min 48.7\Sec & Scd & 8872 & 15.52 & -  \\

       d & 21h 17m 31.2s & -04\Deg 07\Min 14.6\Sec & Sm & 8857  & 16.27  & - \\

         &  &  &  &  &  &   \\

HCG 100 a & 23h 58m 46.3s & +12\Deg 49\Min 57.2\Sec & Sb & 5300 &
13.66 & 30 \\

        b & 23h 58m 52.4s & +12\Deg 50\Min 03.8\Sec & Sm & 5253 & 14.90 & 23 \\

        c & 23h 58m 39.8s & +12\Deg 51\Min 56.1\Sec & SBc & 5461 & 15.22 & 26  \\

        d & 23h 58m 41.0s & +12\Deg 50\Min 03.3\Sec & Scd & - &  15.97 & - \\

\enddata

\tablenotetext{1}{Optical systemic velocities from Hickson 1993}


\tablenotetext{2}{From NED}

\end{deluxetable}

\clearpage

%
%

\begin{deluxetable}{rcccccccc}

\tabletypesize{\scriptsize}

\tablewidth{0pt}
\tablenum{4}
\tablecolumns{9}
\tablecaption{Kinematical Properties of the HCGs}

\tablehead{
\colhead{Name} & \multicolumn{3}{c}{Position Angle (in degree)} & \multicolumn{3}{c}{Inclination (in degrees)}& \colhead{Rotation velocity}  & \colhead{Systemic velocity}\tablenotemark{1}\\
\colhead{} & \colhead{} & \colhead{} & \colhead{} & \colhead{} & \colhead{} & \colhead{} & \colhead{}\\
\colhead{} & \colhead{Velocity} & \colhead{Cont.} & \colhead{Mono.} & \colhead{Velocity} & \colhead{Cont.} & \colhead{Mono.} & \colhead{Approaching / Receding side} & \colhead{} \\
\colhead{} & \colhead{map} & \colhead{map} & \colhead{map } &
\colhead{map} & \colhead{map} & \colhead{map} & \colhead{velocity
($km~s^{-1}$)} & \colhead{$km~s^{-1}$}}

\startdata

HCG 88 a & 128 $\pm$5 & 126 $\pm$10 & 120 $\pm$10 & 65 $\pm$5 & 55
$\pm$3 & 67 $\pm$5 & -280 / +300 &  5963\\

       b & 160 $\pm$5  & 34 $\pm$10 & -     & 55 $\pm$3 & 45 $\pm$? & - & -260 / +265 &  6024 \\

       c & 150 $\pm$3  & -      & -     & 42 $\pm$2 & - & - & -121 / +130 & 5979 \\

       d & 70  $\pm$5  & 70 $\pm$8   & 70 $\pm$ 8 & 85 $\pm$2 & 73 $\pm$3 & 73 $\pm$5  & -200 / +220 &  6037 \\

         &  &  &  &  &  &   &   & \\

HCG 89 a & 54 $\pm$6 & 44 $\pm$10 & 44 $\pm$10 & 45 $\pm$5 & 51
$\pm$3 & 51 $\pm$5 & -214 /  +220 & 8736 \\

       b & 165 $\pm$5 & 155 $\pm$10 &  & 49$\pm$3 & 65$\pm$3 & - & -165 / +130 &  8778 \\

       c &  0 $\pm$3 & 0 $\pm$3 & 0 $\pm$5 & 46 $\pm$3 & 65 $\pm$3 & 65 $\pm$5 & -165 / +190 & 8799 \\

       d & 70 $\pm$7 & 87 $\pm$7 & 85 $\pm$8 & 50 $\pm$5 & 45 $\pm$3 & 45 $\pm$5 & -55 / +65 & 8775 \\

       x & 80 $\pm$3 & 90 $\pm$3 & 90 $\pm$3 & 50 $\pm$3 & 30 $\pm$3 & 30 $\pm$5 & -78 / +62 & - \\

         &  &  &  &  &  &   &  &  \\

HCG 100 a & 78 $\pm$5 & 78 $\pm$5 & 78 $\pm$10 & 50 $\pm$7 & 47 $\pm$3 & 59 $\pm$5 & -190 / +190 &  5323 \\

        b & 145 $\pm$5 & 165 $\pm$5 & 10 $\pm$5 & 52 $\pm$10 & 57 $\pm$3 & 55 $\pm$5 & +13 / +170 & 5163 \\

        c & 72 $\pm$8  & 70 $\pm$5 & 50 $\pm$5  & 66 $\pm$5  & 65 $\pm$3 & 48 $\pm$5 & -70 / +100 & 5418 \\

        d & 53 $\pm$3  & 47 $\pm$5 & 45 $\pm$5 & 85 $\pm$5  & 64 $\pm$3 & 64 $\pm$5  & -170 / +240  & -\\

        x & 100 $\pm$8 & 90 $\pm$8 & 100 $\pm$8 & 50 $\pm$10& 44$\pm$10 & 38$\pm$10 & -150 / +10 & \\

\enddata

\tablenotetext{1}{in $km~s^{-1}$, from our velocity field and rotation curve}

\end{deluxetable}

%
%

\begin{deluxetable}{lcccc}

\tabletypesize{\scriptsize}

\tablenum{5}

\tablecolumns{5}
\tablewidth{0pt}
\tablecaption{Interaction
Indicators}

\tablehead{ \colhead{Indicators} & \colhead{H88a} &
\colhead{HGC 88b}& \colhead{HGC 88c}& \colhead{HGC 88d}}

\startdata
Highly disturbed velocity field & - & + & - & - \\
Central double nuclei & - & - & - & - \\
Double kinematic gas component & - & - & - & -\\
Changing PA along major axis & - & + & - & - \\
Gaseous versus stellar major-axis misalignment & - & + & - & - \\
Tidal tails & - & - & - & -\\
High IR luminosity & + & + & - & -\\
Central activity & + & + & - & -\\
 \hline
 \hline
                & HGC 89a & HGC 89b & HGC 89c & HGC 89d \\
\hline
Highly disturbed velocity field & + & + & + & + \\
Central double nuclei & - & - & - & - \\
Double kinematic gas component & - & - & - & - \\
Changing PA along major axis & + & + & - & + \\
Gaseous versus stellar major-axis misalignment & + & + & - & + \\
Tidal tails & - & + & - & - \\
High IR luminosity & - & - & - & - \\
Central activity & - & - & - & -\\
\hline
\hline
                & HGC 100a & HGC 100b & HGC 100c & HGC 100d \\
\hline
Highly disturbed velocity field & + & + & + & + \\
Central double nuclei & - & + & - & - \\
Double kinematic gas component & - & - & - & - \\
Changing PA along major axis & + & + & + & - \\
Gaseous versus stellar major-axis misalignment & - & + & + & - \\
Tidal tails & - & + & - & - \\
High IR luminosity & + & - & - & - \\
Central activity & - & - & - & -\\
\enddata

\end{deluxetable}

\end{document}